\documentclass[sigconf, nonacm]{acmart}

\usepackage{xcolor}
\usepackage{framed}            
\definecolor{shadecolor}{RGB}{245,245,245} 
\usepackage{fancyhdr}
\usepackage{tabularx}
\newif\ifshowrevisions
\showrevisionstrue               

\newif\ifshowrevnotes
\showrevnotestrue

\newif\ifshowrevmarks
\showrevmarkstrue

\usepackage{soul} 
\usepackage{color}
\usepackage[ruled,vlined,linesnumbered]{algorithm2e} 
\SetKw{KwRet}{return}
\usepackage{amsmath}

\usepackage{amssymb}
\usepackage[normalem]{ulem}
\usepackage{subcaption} 
\usepackage{array} 
\usepackage{multirow} 
\usepackage[most]{tcolorbox}
\usepackage{enumitem}

\newtcolorbox{promptbox}[1]{
    colback=gray!5!white,
    colframe=gray!75!black,
    fonttitle=\bfseries,
    title=#1,
    boxrule=0.5pt,
    arc=2pt,
    left=5pt,
    right=5pt,
    top=5pt,
    bottom=5pt,
    breakable 
}
\usepackage[textsize=footnotesize]{todonotes}  

\newcommand{\Ts}{\mathcal{T}^*}  

\newcommand{\T}{\mathcal{T}}  
\newcommand{\Q}{\mathcal{Q}}  
\newcommand{\sys}{\texttt{EcoTable}\ }  

\newcommand\vldbdoi{XX.XX/XXX.XX}

\newcommand\vldbavailabilityurl{URL_TO_YOUR_ARTIFACTS}

\definecolor{deepgreen}{RGB}{0,100,0}  

\newcommand{\ROOTH}[1]{\textcolor{blue}{#1}}

\allowdisplaybreaks
\begin{document}
\twocolumn                     
\pagenumbering{roman}      
\setcounter{page}{1}       
\pagestyle{plain}          

\setcounter{table}{0}                        
\renewcommand{\thetable}{\Alph{table}}       
\renewcommand{\theHtable}{Response.\thetable}
\setcounter{equation}{0}                         
\renewcommand{\theequation}{R\arabic{equation}}  
\renewcommand{\theHequation}{Response.\theequation} 
\setcounter{figure}{0}                         
\renewcommand{\thefigure}{O\arabic{figure}}    
\renewcommand{\theHfigure}{Response.\thefigure} 


\pagenumbering{arabic}     
\setcounter{page}{1}       

\setcounter{table}{0}                    
\renewcommand{\thetable}{\arabic{table}}  
\renewcommand{\theHtable}{\thetable}      

\setcounter{equation}{0}                     
\renewcommand{\theequation}{\arabic{equation}}  
\renewcommand{\theHequation}{\theequation}      
\setcounter{figure}{0}                      
\renewcommand{\thefigure}{\arabic{figure}}  
\renewcommand{\theHfigure}{\thefigure}      

\setcounter{algocf}{0} 
\renewcommand{\thealgocf}{\arabic{algocf}}
\renewcommand{\theHalgocf}{\thealgocf}

\title{EcoTable: Cost-effective Table Integration in Data Lakes for Natural Language Queries}

\author[]{Yuhui Wang, Jinqi Liu, Chengliang Chai, Hangyu Zhao, Yuhao Deng,\\Yuyu Luo, Xin Tang, Ye Yuan, Guoren Wang, Fengjin Wang$^*$, Lei Cao$^{\dagger}$}
\affiliation{
    \institution{University of Arizona$^{\dagger}$, Kuaishou Technology$^*$, Beijing Institute of Technology}
    \country{USA, China}
}

\begin{abstract}
The diverse formats of CSV and Parquet files in data lakes pose a significant challenge to traditional ETL, which relies on data engineers to pre-define a target database schema and build a complex pipeline for data integration.
Moreover, with this approach, the integrated data often cannot support various analytical needs, as the predefined schema does not necessarily satisfy the table format or join relationships required to answer unforeseen queries.
To address this, we propose \sys, the first natural language-based data integration framework. Given a set of user-specified natural language queries, \sys automatically integrates the tables into a form that adequately supports the corresponding SQL queries. \sys achieves this by leveraging the semantic understanding and complex reasoning capabilities of Large Language Models (LLMs). Moreover, \sys addresses the scalability and cost issues introduced by expensive LLM inferences with a set of novel ideas.

\sys first introduces a graph to represent the overall search space, where nodes represent tables and edges carry weights indicating join likelihood produced by a lightweight deep learning model. On top of this graph data structure, \sys designs three components to achieve our goal: (1) the table identification layer aims to identify relevant tables via a two-stage schema linking based on user queries; (2) the graph-based validation layer aims to discover significant join paths, including necessary data transformations and bridging tables, by modeling the problem as Steiner tree searches; and (3) the table transformation layer generates transformation code to implement the joins using LLMs. 
We construct 4 real-world benchmark datasets with more than 200 queries. Extensive experiments demonstrate that \sys outperforms the state-of-the-art baselines, increasing accuracy by more than 30\% and cutting LLM invocation costs by 5 times.
\end{abstract}

\maketitle



\section{INTRODUCTION}\label{sec:intro}
\noindent \textbf{Motivation.} Data lakes store large collections of files (e.g., in CSV or Parquet formats) that do not fit neatly into relational tables, offering immense potential for data-driven decision making~\cite{datalake_survey,datalake_survey2}. However, deriving values from such a lightly structured data lake is notoriously difficult. The predominant approach relies on a traditional Extract-Transform-Load (ETL) process, where data engineers first \textit{manually} define a target schema and then develop a complex pipeline to integrate data into a data warehouse ~\cite{auto-pipeline,FlowETL}. This is challenging. First, it is rather difficult to pre-define an appropriate database schema that meets the diverse analytical needs of an application. Second, even if the schema were defined correctly, it is labor intensive to convert the data lake tables into the desired relational table format.

Next, we use a real example to better illustrate this.

\begin{figure*}[t]
    \centering
    \includegraphics[width=1\linewidth]{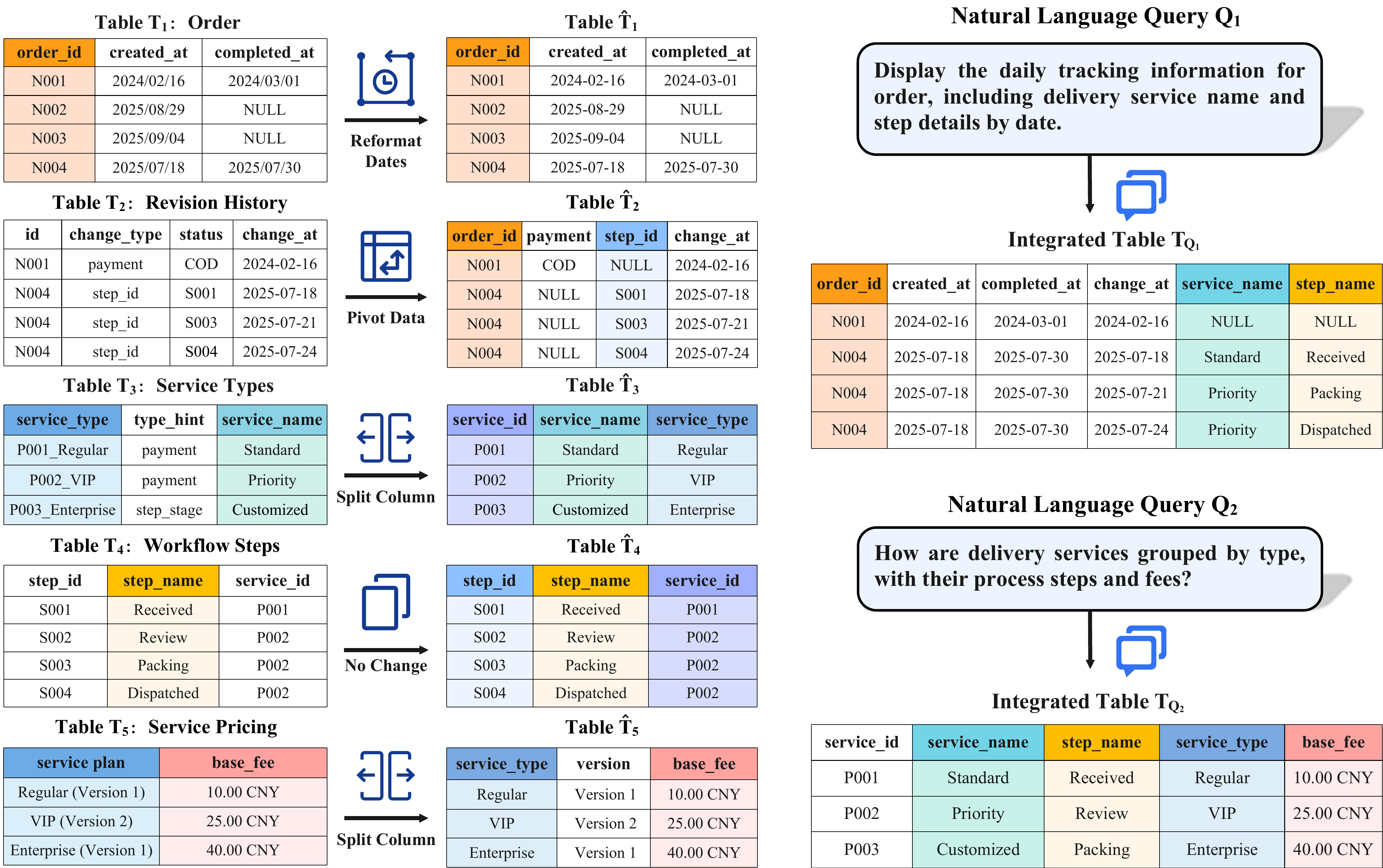}
    \vspace{-2em}
    \caption{An Example of Query-driven Table Integration in Data Lakes.}
    \label{fig:example}
    \vspace{-1.5em}
\end{figure*}

\begin{example}
\label{example 1}
As shown in Figure~\ref{fig:example}, consider 5 tables (i.e., $T_1$: \texttt{Order}, $T_2$: \texttt{Revision History}, $T_3$: \texttt{Service Types}, $T_4$: \texttt{Workflow Steps}, and $T_5$: \texttt{Service Pricing}) sampled from a real-world E-Commerce data lake. A data engineer might naturally define a database schema over these tables, where $T_1$.\texttt{order\_id} joins with $T_2$.\texttt{id}.
However, the pre-defined schema does not necessarily satisfy the analytical need of users. For instance, suppose a user wants to check order status and thus issues a query $Q_1$, \textit{``Display the daily tracking information for the order, including the name of the delivery service and workflow step details by date.''} 
This query involves three tables (i.e., $T_1$: \texttt{Order}, $T_3$: \texttt{Service Types}, and $T_4$: \texttt{Workflow Steps}). However, in the initial database schema, it is impossible to join these tables to answer this query because $T_1$ lacks explicit join conditions with either $T_3$ or $T_4$.
Furthermore, integrating tables requires significant manual efforts. To join $T_1$, $T_3$ and $T_4$ for $Q_1$, a data engineer must: (1) bridge $T_1$ and $T_4$ via $T_2$ with the join between $T_2$ and $T_4$ on the shared \texttt{step\_id} attribute; and (2) transform $T_3$ by splitting \texttt{service\_type} to enable join with $T_4$ on \texttt{service\_id}. 
This process is complex, as it necessitates incorporating additional tables into join path and transforming tables. 
Moreover, the user might want to know more about service fees and thus submit another query $Q_2$: \textit{``How are delivery services grouped by type, with their process steps and fees?''}. To answer $Q_2$, the data engineer must integrate $T_5$: \texttt{Service Pricing} with $T_3$ and $T_4$ -- a new data integration requirement. The key step is to split the \texttt{service\_plan} column in $T_5$ (e.g., ``Regular (Version 1)'') into distinct \texttt{service\_type} and \texttt{version} columns. This prepares $T_5$ to be joinable with $T_3$ on \texttt{service\_type} column. 

This example shows that different queries have different requirements on the database schema. Therefore, pre-defining a database schema to meet users' analytical needs tends to be impractical. Instead, data integration has to be tailored to user queries.
\end{example}





Targeting this problem, we propose to build an automatic table integration system, called \sys. With this system, users only have to provide a set of natural language (NL) queries to intuitively describe their analysis intention. The system will automatically discover a database schema that adequately supports these queries and then perform data integration accordingly.

Existing work has proposed techniques to automate data integration such as schema matching~\cite{magneto, tablegpt, two-step-match}, entity resolution~\cite{In-Context-entity-resolution,em-plm,entity_resolution,DBLP:conf/icde/ShahbaziWMB24,DBLP:conf/deem/Wang022,DBLP:conf/cikm/Wang0H21}, data transformation~\cite{tabletransform,tabulax,DBLP:journals/tkde/ZhangWWX20}, discovery of joinable columns~\cite{deepjoin,fuzzy-integration}, etc. However, all of these approaches rely on a pre-defined target schema. They do not solve the problem of automatically discovering an appropriate database schema based on NL queries and tables. On the other hand, a line of research~\cite{auto-tables, auto-pipeline, auto-suggest,auto-prep} focuses on transforming  tables, e.g., spreadsheets or web tables, into relational formats. However, because the relational schema these methods automatically discover is blind to user queries, it tends to be ineffective in supporting the analytical needs of users, as shown in the above example.



The semantic understanding and reasoning ability of Large Language Models (LLMs) make it possible for us to achieve the above goal. First, existing Text-to-SQL research has shown that LLMs excel at identifying the mapping between NL queries and columns in the tables through schema linking~\cite{chess,macsql,reforce,RSLsql}.   Moreover, with its reasoning ability ~\cite{chain-of-table, plan-and-solve,react, spider2}, an LLM is able to produce a plan that combines multiple operations to collaboratively solve the data transformation problem.

\noindent \textbf{Challenge.} 
An intuitive method that leverages LLMs to solve this problem is to ask LLMs to (1) identify all tables that are relevant to the NL queries provided by users; (2) check whether every pair of these discover tables is joinable, thus finding a join path for each query; (3) finally, perform data transformation and integration accordingly.
However, routinely invoking LLMs to explore all possible join relationships is prohibitively expensive. This is because the number of potential join paths grows exponentially with the number of tables. Therefore, the key challenge lies in how to minimize the number of LLM calls while still achieving high-quality integration to reliably answer user queries.

\noindent \textbf{Our Proposal.} To address this challenge, we propose \texttt{EcoTable}, a cost-effective query-driven table integration system. The key idea is to combine smaller deep learning models and LLMs to efficiently identify and validate potential join paths. At a high level, deep learning models give prior probabilities of possible join relationships between tables. Then, to minimize LLM invocations, \sys prioritizes the pairs of tables for the LLMs to validate that are (1) most likely to be joinable and (2) required to answer user queries.

\sys models the search space as a fully connected graph where each node corresponds to a table, and each edge (i.e., the join relationship) connects two tables with a weight indicating the join likelihood initialized with a deep learning model. 
In this way, the problem of finding a {\it valid join path} for each NL query is converted to identify a connected subgraph that covers the tables involved in the query.
%
\sys consists of three key components that efficiently discover valid join paths and perform data transformation on top of the graph.

\noindent\underline{\textit{Table Identification Layer}}
identifies query-related tables (i.e., nodes) from all data lake tables via a two-stage schema linking method that first leverages a lightweight deep learning model for coarse-grained filtering, followed by using LLMs for fine-grained verification. This step effectively reduces the search space to a subset of nodes in the above graph, ensuring that subsequent steps primarily focus on these relevant tables.

\noindent\underline{\textit{Graph-based Validation Layer.}}
Next, \sys discovers valid join paths to integrate these tables. As observed in the above example, joining these tables may require additional bridging tables, which is challenging to discover among a potentially large number of tables. To address this, we model the problem as a Steiner tree search problem, where the goal is to find the minimal-weight tree, i.e., a tree composed of edges with the highest join likelihood, which connects all relevant tables for each query, potential via a set of bridging tables.

More specifically, \sys gathers edges from Steiner trees w.r.t. various queries and submits them to LLMs for batch verification. This opens the opportunity to leverage shared edges among queries to reduce LLM costs. Then \sys updates the edge weights of the graph based on verified results and iteratively searches the Steiner trees until the necessary edges of all queries have been validated. In this way, \sys selectively applies LLMs to validate the most promising join relationships and progressively discovers valid join paths with minimal LLM usage.

\noindent\underline{\textit{Table Transformation Layer.}}
\sys then applies the necessary data transformation operations to perform joins and integrate tables. \sys uses LLMs to automatically generate transformation code for each valid join path, guided by a ReAct-style reasoning process that alternates between transformation and join validation.  To reduce latency, \sys further employs a graph-based parallelization strategy, which identifies parallelizable transformations by modeling the problem as an edge coloring task on the validated join paths.

In summary, we make the following contributions.

\noindent(1) We propose \texttt{EcoTable}, to the best of our knowledge, the first natural language driven system for table integration that minimizes LLM invocations while adequately supporting user queries. 

\noindent(2) We design a graph-based validation layer that models table relationships as a weighted graph and identifies valid join paths using a Steiner tree–based optimization.

\noindent(3) We introduce a table transformation layer that employs a ReAct-style reasoning loop to automatically generate transformation code for joinable tables and a graph-based parallelization strategy to further reduce latency and LLM invocation cost.

\noindent(4) We construct 4 real-world benchmark datasets covering multiple domains with more than 200 natural language queries. Comprehensive experiments show that \sys surpasses state-of-the-art baselines by achieving over 30\% accuracy improvement and reducing LLM invocation costs by 5 times.

\begin{figure*}[t]
    \centering
    \includegraphics[width=0.80\linewidth]{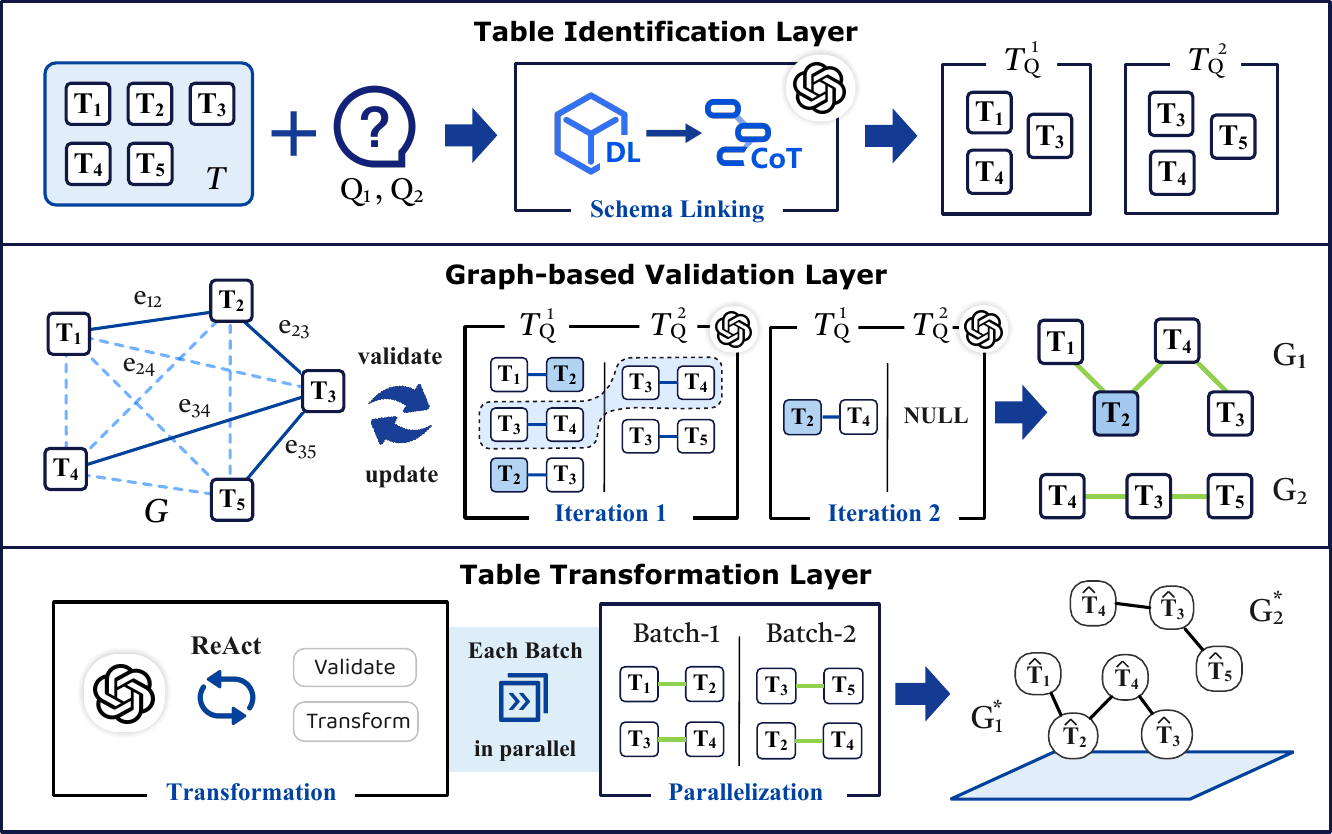}
    \vspace{-1em}
    \caption{The Overall Framework of \texttt{EcoTable}.}
    \vspace{-1.5em}
    \label{fig:autointegrate}
\end{figure*}

\section{PROBLEM DEFINITION}
Given a data lake including a large number of tables (e.g., CSV or parquet files) and a set of user-specified natural language queries, we aim to integrate subsets of tables related to these queries such that each query could be well addressed by an integrated table. 
Due to the heterogeneity of the data lake tables, the integration process might be rather complicated, including transformations and joins over multiple tables. Formally, we define the problem as follows.

\begin{definition}[Query-Driven Data Integration]~\label{def:problem}
Given $N$ tables in a data lake $\T = \{T_1, T_2, \dots, T_N\}$ and a set of $M$ natural language queries 
$\Q = \{Q_1, Q_2, \dots, Q_M\}$, 
we aim to construct a set of transformed tables 
$\Ts = \{\hat{T}_1, \hat{T}_2, \dots, \hat{T}_n\}$ from $\T$ 
to form a global join graph 
$\mathcal{G}^{*} = (\Ts, E^{*})$, 
such that each query $Q_k \in \Q$ 
can be successfully answered by querying a connected subgraph 
$\mathcal{G}^{*}_k = (\Ts_k, E^{*}_k)$ of $\mathcal{G}^{*}$, where $\Ts = \bigcup_{k=1}^{M} \Ts_k$ and $E^{*} = \bigcup_{k=1}^{M} E^{*}_k, k \in [1, M]$.
That is, each connected subgraph $\mathcal{G}^{*}_k$ corresponds to an integrated table sufficient for answering query $Q_k$ and the table is obtained by integrating tables in $\Ts_k$.
The nodes and edges are defined as follows:

\paragraph{\textbf{Nodes:}} 
    For each query $Q_k$, the node set $\Ts_k = \{\hat{T}_{k}^1, \hat{T}_{k}^2, \dots \}$ contains the transformed tables required to answer the query. Each $\hat{T}_{k}^i \in \Ts_k$ is derived from a source table $T_{k}^i \in \T_k \subseteq \T$ through the transformation $\phi_{k}^i$, i.e., 
    $\hat{T}_{k}^i = \phi_{k}^i(T_{k}^i)$.
    In this paper, $\phi_{k}^i(\cdot)$ includes operations such as reformatting, splitting, and pivoting.

\paragraph{\textbf{Edges:}}
    For any edge $\hat{e}^{ij}_{k} \in E^{*}_k$, it connects two tables 
    $\hat{T}^{i}_{k}, \hat{T}^{j}_{k} \in \Ts_k$ 
    if they can be joined to answer $Q_k$. 
    Note that $\mathcal{G}^{*}_k$ is connected, corresponding to the join path, but $\mathcal{G}^{*}$ might not be when different queries are associated with totally different tables.

\end{definition}

\begin{example} \label{ex:formal_def}
Based on Example \ref{example 1} and Figure \ref{fig:example}, the mapping to the formal graph $\mathcal{G}^*$ is as follows:
\begin{itemize}
    \item Query $Q_1$ is addressed by subgraph $\mathcal{G}_{1}^*$, with nodes $\T_1^* = \{\hat{T}_1, \hat{T}_2, \hat{T}_3, \hat{T}_4\}$ and edges $E_1^* = \{\hat{e}_{1}^{12}, \hat{e}_{1}^{24}, \hat{e}_{1}^{34}\}$. Each $\hat{T}_i$ is the result of the transformation $\phi_{1}^i$ (e.g., date reformatting for $T_1$, table pivoting for $T_2$, and column splitting for $T_3$).
    \item Query $Q_2$ is addressed by subgraph $\mathcal{G}_2^*$, which involves nodes $\T_2^* = \{\hat{T}_3, \hat{T}_4, \hat{T}_5\}$ and edges $E_2^* = \{\hat{e}_2^{12}, \hat{e}_2^{13}\}$. Here, $\hat{T}_5$ is the result of the transformation $\phi_2^3$ (e.g., column splitting for $T_5$).
    \item The global join graph $\mathcal{G}^* = (\T^*, E^*)$ is defined by the unions $\T^* = \bigcup_{k=1}^M \T_k^*$ and $E^* = \bigcup_{k=1}^M E_k^*$, forming a unified structure that supports all analytical requirements.
\end{itemize}
\end{example}

\noindent {\textbf{Table Transformation}} resolves structural and semantic discrepancies in data lakes.We focus on three essential operators:

    \noindent \underline{\textit{Normalize}} standardizes data formats for consistency, e.g., converting the \texttt{created\_at} date in Table $T_1$ from "2024/02/16" to the ISO standard "2024-02-16" in Table $\hat{T}_1$.
    
    \noindent \underline{\textit{Split}} decomposes composite columns into atomic ones, e.g., splitting the \texttt{service\_type} column in Table $T_5$ (e.g., "Regular (Version 1)") into separate \texttt{service\_type} and \texttt{version} columns in Table $\hat{T}_5$.
    
    \noindent \underline{\textit{Pivot}} restructures tables by transposing unique values into new columns, e.g., applying \textit{Pivot} to  $T_2$ on the \texttt{change\_type} column generates new column headers  \texttt{payment} and \texttt{step\_id} in Table $\hat{T}_2$.

\noindent {\textbf{Join Discovery}}
identifies semantically equivalent columns across tables to form valid join paths. Existing approaches include:

\noindent \underline{\textit{Deep learning methods}} (e.g., DeepJoin~\cite{deepjoin} and OmniMatch~\cite{omnimatch}) encode column profiles into vector representations and predict the join likelihood between columns using PLMs. For example, $T_2$.\texttt{change\_type} and $T_3.\texttt{type\_hint}$ might be considered as joinable because PLM-based methods focus more on token overlaps or string similarities. In reality, they refer to different things: \texttt{change\_type} is a revision category, whereas \texttt{type\_hint} is an interface tag (e.g., \texttt{step\_stage}) denoting the record’s current workflow stage, so they should not be joined.
 
\noindent \underline{\textit{LLM-based methods}} leverage fine-tuning ~\cite{tablegpt} or prompt engineering~\cite{magneto} to understand table semantics and validate join relationships. These methods demonstrate strong few-shot learning capabilities and can handle complex semantic matching scenarios that are challenging for traditional approaches. By reasoning over column descriptions and data patterns, LLMs achieve robust performance even on unseen schema or ambiguous relationships.

Given $\T$ and $\Q$, a straightforward way to construct $\mathcal{G}^*$ is that for each query $Q_k$, we enumerate a large number of possible table combinations from $\T$ and identify the combination that can satisfy the query, which is  prohibitively expensive if we have to call LLMs to verify whether each combination is valid. Thus, in this paper, we propose the \sys framework to largely reduce the LLMs costs.


\section{OVERVIEW}
\label{sec.overview}


As illustrated in Sec.~\ref{sec:intro}, \sys formulates the query-driven table integration task as a join path identification and validation problem over a graph, ensuring a balance between integration effectiveness and LLM invocation cost. Specifically, \sys first constructs a fully connected graph $\mathcal{G} = (\mathcal{T}, E)$. 
Then, \sys leverages the following three layers to integrate tables for correctly answering the given queries. As shown in Figure~\ref{fig:autointegrate}, the first  table identification layer takes the queries as input, and outputs the table subsets relevant to these queries. Second, the graph-based validation layer takes these tables as input, and outputs valid join paths  through searching within the graph $\mathcal{G}$, so as to connect these relevant tables. Third, the table transformation layer takes as input these identified join paths, applies necessary data transformation operations to actually join each two tables and outputs the integrated tables. 

%
%


\subsection{Table Identification Layer}
\label{sec:3.1}
In Figure ~\ref{fig:autointegrate}, for example, given two queries $Q_1$ and $Q_2$, the module applies schema linking to obtain the corresponding relevant table sets $\mathcal{T}_Q^1 = \{T_1, T_3, T_4\}$ and $\mathcal{T}_Q^2 = \{T_3, T_4, T_5\}$, a well-studied process~\cite{dcgsql,dinsql,RSLsql}. To strike a balance between accuracy and LLMs cost for identifying these tables, we adopt a hybrid approach of combining LLMs and deep learning models.
Specifically, we first train a deep learning model $M_s$ to predict the relevance between an NL query and a table. This model takes as input the concatenation of the query and the table's column names (i.e., its schema), embeds them and outputs a relevance score. Tables with high scores are selected as candidates for subsequent verification by LLMs (see Sec. 4.1 for details).

\subsection{Graph-based Validation Layer}
\label{sec3.2}
As shown in Figure~\ref{fig:autointegrate}, given the identified table subsets $\mathcal{T}_Q^1$ and $\mathcal{T}_Q^2$, this module finds valid join paths $\mathcal{G}_1$ and $\mathcal{G}_2$ that connect them.
As outlined in Sec.~\ref{sec:intro}, identifying such join paths is challenging because: (1) making relevant tables joinable often requires appropriate data transformations or added bridge tables, and (2) the search space is large to identify the valid join path.

To address this, we formulate it as a Steiner tree search problem based on the fully connected graph $\mathcal{G}$.
Initially, \sys uses a deep learning model $M_J$ to estimate the likelihood of joining pairs of tables and assign these values as edge weights in the graph.
%
%
After that, for each $Q_k$, \sys attempts to search a corresponding Steiner tree from $\mathcal{G}$, where (1) the nodes cover the tables in $\T_Q^k$ and (2) the edges have the maximum overall joinable probability. 
Naturally, this Steiner tree represents the most promising join path in which the tables in $\T_Q^k$ could be joined to answer $Q_k$ (e.g., for $\mathcal{T}_Q^1 = \{T_1, T_3, T_4\}$, a Steiner tree introduces $T_2$ as a bridge with edges $\{e_{12},e_{23}, e_{34}\}$, as shown in Figure~\ref{fig:autointegrate}).

However, deep learning is typically not accurate enough, so we collect all edges from the Steiner trees corresponding to different queries and employ LLMs for further verification.
This processing leverages the fact that multiple queries may share common edges, thereby saving LLMs costs (e.g., edge $e_{34}$ appears in both $\mathcal{T}_Q^1$ and $\mathcal{T}_Q^2$'s Steiner trees and requires only LLM validation once). Afterwards, we update the graph by removing edges that are deemed unjoinable by LLMs, while assigning a weight of one to those edges that are confirmed as joinable. We then iteratively recompute the Steiner trees using the updated graph and iterate the validation process until for every query, we either find a join path that makes the query executable or conclude that no such path exists due to insufficient connectivity (see Section 4.2 for details).

\subsection{Table Transformation Layer}
\label{sec3.3}
Finally, in Figure~\ref{fig:autointegrate}, starting from the valid join paths $\mathcal{G}_1$ and $\mathcal{G}_2$, the module applies data transformation operations to get $\mathcal{G}_1^*$ and $\mathcal{G}_2^*$, the integrated tables used to answer queries $Q_1$ and $Q_2$.
To be specific, for each valid edge, \sys uses a transformation modular component $M_T$ to automatically produce executable code, defined in a formal transformation grammar, for the two tables connected by the edge.
To improve accuracy, LLMs verify each transformation using a ReAct-style reasoning process. 
For all these edges, a straightforward method is to batch multiple tables and ask LLMs to transform them together. However, this would degrade the accuracy due to complex chain dependencies within these tables, which is difficult for LLMs to understand. An alternative is to sequentially process each edge. However, this would incur high latency due to the iterative process of ReAct.
To tackle this issue, \sys further introduces a graph-based parallelization strategy to identify independent transformations that can be executed concurrently based on the edge-coloring framework. The key observation is that two edges can be transformed in parallel if they do not share the same table.  For example, in $\mathcal{G}_1$, transformations for $e_{12}$ ($T_1$-$T_2$) and $e_{34}$ ($T_3$-$T_4$) can be processed in parallel (see Section 5 for details).
%


\section{GRAPH-BASED TABLE INTEGRATION}~\label{sec:graph}
In this section, we introduce in details how \sys identifies  query-related tables (Section~4.1), searches valid join paths (Section~4.2) and leverages LLMs to verify promising join relationships to produce finally integrated tables (Section~\ref{sec:validation}).

\subsection{Table Identification Layer} \label{sec:identify} 
\noindent \textbf{Stage 1 (PLM Table-level Selection).} 
We fine-tune a lightweightcompact pre-trained language model, $M_S$, as a cross-encoder to compute relevance scores between a query and a table schema. Unlike bi-encoder setups that embed queries and schemas separately, $M_S$ encodes them together in one sequence, allowing full-attention interactions between the query and schema metadata. As a modular component, $M_S$ can be instantiated flexibly with various pre-trained backbones; in our implementation, we specifically support RoBERTa~\cite{roberta}, RoBERTa-Large~\cite{debertav3}, and DeBERTa-v3-Large~\cite{XLM-RoBERTa-XL}. Subsequently, for each query, i.e., $Q_k \in \mathcal{Q}$, we choose the tables with large relevance scores as candidates, denoted as $\T_\text{cand}^k$.
To be specific, we train the model on inputs of the form ``\textit{query} [TAB] \textit{table\_name} [COL] \textit{col}\_1 [COL] \textit{col}\_2 [COL], ...''. The training process optimizes the model to predict binary labels $y_i \in \{0, 1\}$ by minimizing cross-entropy loss:

\begin{equation}
     \mathcal{L} = -\left[y_i \log s_i + (1-y_i) \log(1-s_i)\right]
\end{equation}

\noindent where $s_i$ denotes the relevance score predicted by the model. 

Overall, this step retrieves a sufficient number of tables that may answer the queries, i.e., ensuring a high recall. However, exploring the join pairs among this relatively large number of tables is still expensive. 

\noindent \textbf{Stage 2 (LLMs Refinement).}
To address this, we employ LLMs to refine those candidate tables. Given $\T_\text{cand}^k$, \sys performs a fine-grained verification using the CoT reasoning capability of LLMs. Specifically, \sys prompts the LLMs to reason through the following steps:

\noindent\underline{\textit{Query Understanding.}}  We utilize the LLMs as a query decomposer following a one-shot prompt setting. Given a natural language query $Q$, the LLMs are guided to split it into a sequence of sub-questions $q_1, q_2, \dots, q_n$, effectively simplifying the complex request into atomic semantic components such as entities and attributes. For example, for $Q_1$ in Figure ~\ref{fig:example}, LLMs identify key entities as \textit{``order''} and \textit{``delivery service''}, attributes such as \textit{``order\_id''}, \textit{``service\_name''}, and \textit{``step\_name''}. This decomposition captures what the query focuses on. 

\noindent\underline{\textit{Schema Reasoning and Alignment.}} For $\T_\text{cand}^k$, the LLMs use Chain-of-Thought to evaluate each candidate table step by step. Specifically, for each decomposed semantic component, the LLM tries to find a matching column in the table. It explicitly explains its reasoning, such as why a column matches a sub-question or why a table is missing the required information. If a candidate table cannot match any necessary components, the LLM rejects it. 
For example, for $Q_1$, LLMs reason that the entity \textit{``order''} from the query is represented by the table $T_1$: \texttt{Order}, while its attribute \textit{``order\_id''} corresponds to the column \texttt{order\_id} in $T_1$. 
In this way, LLMs reliably identify the final verified table set $\T_{Q}^k \subseteq \T_\text{cand}^k$, which contains the tables that are necessary to answer the query.

\subsection{Graph-based Validation Layer}~\label{sec:validation}

Given the fully-connected global join graph $\mathcal{G}=(\mathcal{T},E)$, each edge $e_{ij}\in E$ represents a potential join relationship between tables $T_i$ and $T_j$, associated with a weight that represents the joinable probability $p(e_{ij}) \in [0,1]$ computed by $M_J$.
Given queries $Q = \{Q_1, Q_2,...,Q_M\}$, \sys discover a collection of query-specific join paths $\{\mathcal{G}_1, \mathcal{G}_2, \dots, \mathcal{G}_M\}$, where each $\mathcal{G}_k=(\mathcal{T}_k,E_k)$ connects  tables in $\T_{Q}^k$ required to answer $Q_k$. 
Formally,  this join path discovery problem can be formulated as:

\begin{equation}~\label{eqa:formulation}
\begin{aligned}
\max_{\{x_{ij}^{k}\}} \quad & 
\sum_{k=1}^{M} \prod_{e_{ij} \in E}  p(e_{ij})^{ x_{ij}^{k}} \\
\text{s.t.} \quad &
\forall k,\, \forall T_i \in \T_Q^k: 
\sum_{e_{ij} \in \delta(T_i)} x_{ij}^{k} \ge 1, \\
& 
\forall k,\, \forall S \subset \T'_k,\, 
S \ne \emptyset,\, S \ne \T'_k: 
\sum_{e_{ij} \in \phi(S)} x_{ij}^{k} \ge 1.
\end{aligned}
\end{equation}

\noindent where $x_{ij}^{k} \in \{0,1\}$ is a binary decision variable indicating whether edge $e_{ij}$ is included in the join path of query $Q_k$. $\delta(T_i)$ denotes the set of edges corresponding to node $T_i$. $\T'_k = \{ \ T_i  | \exists T_j, x_{ij}^k = 1\}$, representing  the set of tables (including the bridging tables) selected in the join path for query $Q_k$.
For a subset $S \subseteq \T'_k$, $\phi(S)$ denotes the cut set of $S$ (i.e., the edges connecting $S$ to its complement $\T'_k \setminus S$), defined as $\phi(S) = \{e_{ij} \in E \mid T_i \in S, T_j \in \T'_k \setminus S\}$.
The first constraint ensures that the join path includes all tables in $\T_Q^k$ for each query $Q_k$, while the second enforces connectivity. This optimization maximizes the sum of the overall joinability of each query, which is computed by multiplying the join likelihoods of edges in the query's join path, i.e., $\prod_{e_{ij} \in E}  p(e_{ij})^{ x_{ij}^{k}}$. In this way, we prioritize the edges that are likely to be joined to answer queries for LLMs to validate.

\noindent\textbf{Pairwise Joinability Estimation.}
Following~\cite{autobi}, \sys adopts a deep learning model $M_J$ to compute $p(e_{ij})$. 
This model serves as an efficient but approximate estimator of joinability, providing a prior probability for each edge. 
Specifically, given two columns, $M_J$ takes input table features (e.g., metadata and column embeddings) and outputs join probabilities in (0,1]. Designed as a modular component, $M_J$ allows for flexible instantiation. In this work, we utilize three models: Auto-BI~\cite{autobi}, DeepJoin~\cite{deepjoin}, and OmniMatch~\cite{omnimatch}.
Formally, let $C \in T_i$ and $C' \in T_j$ denote two columns in tables $T_i$ and $T_j$. $M_J(C_i, C_j) \in (0,1]$ predicts the probability that $C$ and $C'$ could be joined. Because the joinability between two tables often depends on the presence of at least one pair of aligned columns, we use the maximum pairwise similarity as a joinability score and assign this as the weight, i.e., $p(e_{ij})=\max_{C \in T_i,\, C' \in T_j} M_J(C, C')$.


\begin{algorithm}[t]
\caption{Graph-based Validation Algorithm}
\label{alg:join-validation}
\DontPrintSemicolon
\KwIn{Global weighted join graph $\mathcal{G}=(\mathcal{T},E)$, relevant tables for each query $\{\T_Q^1, \T_Q^2, ..., \T_Q^M\}$.}
\KwOut{Validated join paths $\{\mathcal{G}_1, \mathcal{G}_2, \dots,\mathcal{G}_M\}$.}
$E^{+} \gets \emptyset,\; E^{-} \gets \emptyset$;\;\nllabel{alg:alg1-1}

\While{$\neg \bigcup_1^k E_k \subseteq E^+ \land \mathrm{HasTree}(\mathcal{G}, \{\T_Q^1,...,\T_Q^M\})$}{\nllabel{alg:alg1-3}
  \textcolor{blue}{\tcp{Step 1: Graph update}}
    $\mathcal{G}_u \gets \mathrm{UpdateGraph}(\mathcal{G}, E^{+}, E^{-})$;\;\nllabel{alg:alg1-4}
  
  \textcolor{blue}{\tcp{Steps 2: LLMs validation}}
  $E_{b} \gets \emptyset$;\;\nllabel{alg:alg1-5}
  \For{$k \in [1..M]$}{\nllabel{alg:alg1-6}
    $\mathcal{G}_k, \T_k, E_k = \mathrm{SteinerTree} (\mathcal{G}_u, \T_Q^k)$;\;\nllabel{alg:alg1-7}
    $E_{b} \gets E_{b} \cup E_k$;\;\nllabel{alg:alg1-9}
  }
  $E_{new} \gets E_{b} \setminus (E^+ \cup E^-)$;\;\nllabel{alg:alg1-10}
  \For{each $e_{ij} \in E_{new}$}{\nllabel{alg:alg1-11}
    \If{$\mathrm{LLMValidate}(e_{ij}) = \text{VALID}$}{\nllabel{alg:alg1-14}
        $E^+ = E^+ \cup \{e_{ij}\}$;\;\nllabel{alg:alg1-15}
    }
    \Else{\nllabel{alg:alg1-16}
        $E^- = E^- \cup \{e_{ij}\}$;\;\nllabel{alg:alg1-17}
    }
  }
}
\For{$k \in [1..M]$}{\nllabel{alg:alg1-18}
    $\mathcal{G}_k = (\T_k, E^+ \cap E_k)$;\;\nllabel{alg:alg1-19}
}
\Return {$\{\mathcal{G}_1, \mathcal{G}_2, \dots, \mathcal{G}_M\}$};\;\nllabel{alg:alg1-20}
\end{algorithm}

\begin{figure*}[t]
    \centering
    \includegraphics[width=1\linewidth]{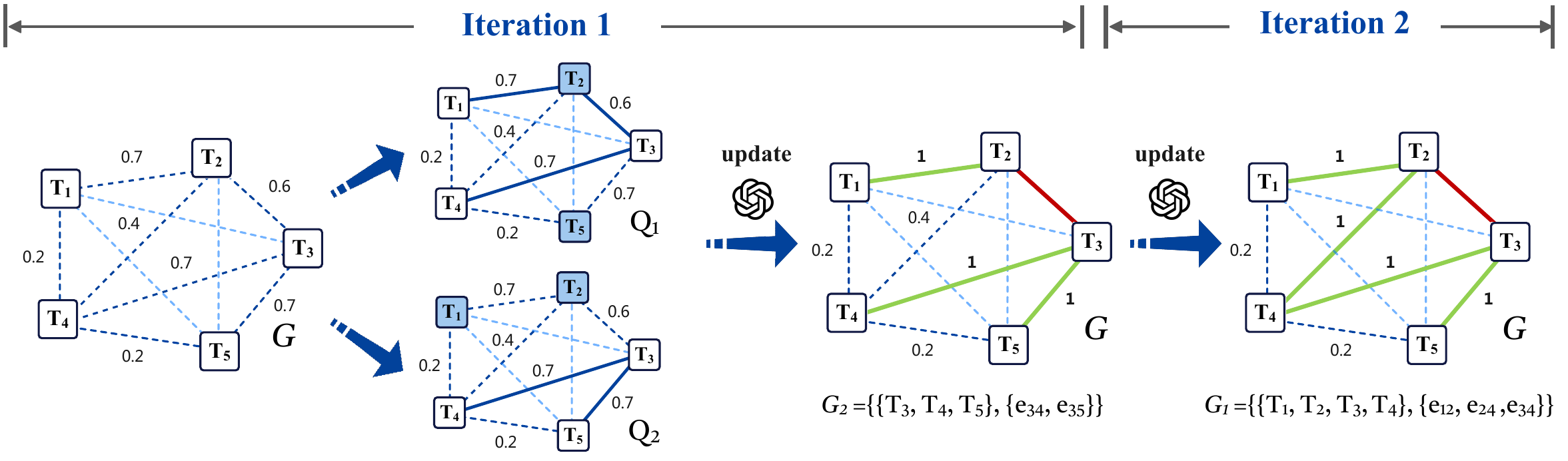}
    \vspace{-2em}
    \caption{An Example of Join Path Search and Validation.}
    \vspace{-1em}
    \label{fig:4.2-Steiner tree}
\end{figure*}

\noindent\textbf{Iterative Validation with LLMs.} 
Based on the weighted join graph $\mathcal{G}$, \sys utilizes the typical algorithm~\cite{kousteiner} with an approximation ratio of 2 (i.e., the KMB algorithm) to efficiently identify a Steiner tree for each query $Q_k$. Then \sys performs iterative validation with LLMs to refine the accuracy of these discovered Steiner trees. Specifically, at the beginning of each iteration, \sys selects all unverified edges from the Steiner tree of each query. These edges are submitted to LLMs to validate join relationship. \sys updates the weight of the  join graph $\mathcal{G}=(\T, E)$ accordingly as follows:

\begin{itemize}
    \item If an edge $e_{ij}$ passes validation, its corresponding weight is set to 1, ensuring that \sys will only validate one edge at most once.
    \item If $e_{ij}$ is invalid, the edge is removed from the graph and will not be considered in later iterations.
\end{itemize}

For the updated graph, \sys recomputes a maximum Steiner tree for each query $Q_k$ and checks the unverified edges in these new trees. This process repeats until all edges of the obtained trees pass the verification, or no more feasible Steiner trees are available.

\noindent\underline{\textit{Remark.}} Since multiple queries often involve overlapping table pairs, \sys caches all validated results globally. If an edge has been validated once—either as valid or invalid—the result is reused directly in the Steiner tree searches of other queries, eliminating redundant LLM calls throughout the query set $\mathcal{Q}$.

\noindent\textbf{Algorithm Overview.} 
Algorithm~\ref{alg:join-validation} summarizes the whole process of the graph-based validation layer. \sys initializes two edge sets $E^+$ and $E^-$ to record valid and invalid joins verified before (line~\ref{alg:alg1-1})
%
In each iteration (line~\ref{alg:alg1-3}), \sys updates the join graph $\mathcal{G}$ by removing invalid edges and setting the weight of valid ones as 1 (line~\ref{alg:alg1-4}). It then searches for a new Steiner tree for each query over the updated graph (line~\ref{alg:alg1-6}-\ref{alg:alg1-7}). Next, the algorithm collects and deduplicates all unverified edges from these new Steiner trees into a candidate set $E_{new}$ (line~\ref{alg:alg1-9}-\ref{alg:alg1-10}).
Each edge is checked by LLMs. \sys adds valid edges to $E^+$ and invalid ones to $E^-$ (line ~\ref{alg:alg1-11}-~\ref{alg:alg1-17}). It then uses the updated graph to recompute new Steiner trees for all queries until all edges pass the validation or no feasible tree remains. Finally, the algorithm outputs the validated join paths (line ~\ref{alg:alg1-18}-~\ref{alg:alg1-20}).

\noindent \underline{\textit{Remark.}} Because LLMs validation is prone to hallucinations, \sys further leverages majority voting for two types of critical edges: (1) edges that appear in Steiner trees corresponding to multiple queries within the same iteration, and (2) edges whose deletion would prevent constructing a new Steiner tree for a query in the subsequent iteration. \sys makes three independent LLM calls for each such edge and then adopts the majority decision.

\begin{example}
As shown in Figure~\ref{fig:4.2-Steiner tree}, at the beginning, \sys constructs the initial weighted graph $\mathcal{G} = (\T, E)$, where the edge weights represent the joinability score estimated by model $M_J$. For each query, \sys first searches a Steiner tree that connects all query-relevant tables with the maximum total joinability score. The initial Steiner tree for $Q_1$ includes edges $\{e_{12}, e_{23}, e_{34}\}$, while for $Q_2$, it covers edges $\{e_{34}, e_{35}\}$.
In the first iteration, \sys validates all unverified edges of both trees using LLM reasoning. Since $e_{34}$ appears in both trees, \sys makes three independent LLM calls for it. Following the majority decision, $e_{34}$ is confirmed as valid. Simultaneously, $e_{12}$ and $e_{35}$ are verified, while $e_{23}$ is rejected and removed. All results are stored in the global cache. The graph $\mathcal{G}$ is updated accordingly: valid edges are set to weight 1, while invalid edges such as $e_{23}$ are removed. At this point, $Q_2$ already has a validated join path: $\mathcal{G}_2 =\{\{T_3,T_4,T_5\},\{e_{34}, e_{35}\}\}$.
In the second iteration, \sys recomputes the Steiner tree for $Q_1$ on the updated graph. The new candidate tree contains edges $\{e_{12}, e_{24}, e_{34}\}$. Since $e_{12}$ and $e_{34}$ are already cached as valid, only $e_{24}$ is newly verified. LLMs confirm $e_{24}$ as valid, completing the join path for $Q_1$: $\mathcal{G}_1 = \{\{T_1,T_2,T_3,T_4\},\{e_{12}, e_{24}, e_{34}\}\}$.
\end{example}

\noindent\textbf{Complexity of graph-based validation.} Eq.~\ref{eqa:formulation} reduces to the Minimum Steiner Tree problem, and pruning via Eq.~\ref{eq:spanner_prob} yields the Minimum $t$-Spanner problem; both are NP-hard~\cite{kousteiner,t-spanner}.

\begin{proof}[Proof of Steiner Tree Reduction]
   Considering a special case when $M = 1$, the problem then simplifies to $\max_{\{x_{ij}^{1}\}}  \prod_{e_{ij} \in E} p(e_{ij})^{x_{ij}^{1}}$. Then, we can convert the problem to find a join path $\mathcal{G}_1=(\mathcal{T}_1, E_1)$ containing all required tables $\T^1_Q$ from $\mathcal{G}$:

    \begin{equation}
        \begin{aligned}
         \max_{\{x_{ij}^{1}\}}  \prod_{e_{ij} \in E} p(e_{ij})^{x_{ij}^{1}}
        &\Longleftrightarrow\; \max_{\{x_{ij}^{1}\}}  \ \ln \!\left(\prod_{e_{ij} \in E} p(e_{ij})^{x_{ij}^1}\right)  \\
        &\Longleftrightarrow\;
        \min_{\{x_{ij}^{1}\}} \sum_{e_{ij} \in E} -\ln p(e_{ij})\, x_{ij}^1
        \end{aligned}
    \end{equation}

    Next, we prove that the  minimum Steiner tree problem~\cite{kousteiner} can be reduced to this special case in polynomial time. The minimum Steiner tree problem is defined as follows. Given a weighted graph $\mathcal{G}'=(\mathcal{T}',E')$ and a set of target nodes $R \subseteq \mathcal{T}'$, the goal is to determine a connected join path $\mathcal{G}_w=(\mathcal{T}_w, E_w) \subseteq \mathcal{G}'$ such that $R \subseteq \mathcal{T}_w$ has the minimum total weight $\sum_{e \in E_w} w'(e)$, where $w'(e)$ denotes the weight of edge $e \in E'$. 
    Let $\mathcal{G}' = \mathcal{G}$, $\mathcal{T}' = \mathcal{T}$, $E' = E$, $w'(\cdot) = -\ln{p(\cdot)}$, and $\mathcal{T}_w = \T^1_Q$. Then, our problems share identical objectives and constraints with the classical minimum Steiner tree problem. Because the classical minimum Steiner tree problem is NP-hard, our problem is also NP-hard.
\end{proof}

Moreover, we analyze the Steiner Tree obtained in $\mathcal{G}_s$. We compute Steiner trees using a 2-approximation method (the KMB algorithm). Since $\mathcal{G}_s$ is a $t$-spanner of $\mathcal{G}$, this implies a $2t$-approximation when working in $\mathcal{G}_s$. Consequently, the Steiner trees obtained in $\mathcal{G}_s$ achieves joinability within a factor of $2t$ of the optimal Steiner trees in $\mathcal{G}$.

We also bound the total number of LLM invocations $\mathcal{N}_{LLM}$ for $M$ queries. The lower bound represents an ideal case where only edges truly belonging to the optimal Steiner trees are verified, with no extra or unnecessary checks. In contrast, the upper bound indicates the worst-case  in which \sys might have to check every edge in the pruned subgraph $\mathcal{G}_s$. The number of edges $|E_s|$ provides this upper bound, as established by $t$-spanner theory. Therefore, we have:
\begin{equation}\Omega\left( \left|\bigcup_{k=1}^{M} E_{OST}(\mathcal{T}_Q^k) \right|\right) \le \mathcal{N}_{LLM} \le \tilde{O}\left(f \cdot |\mathcal{T}|\cdot \sum_{k=1}^M|\mathcal{T}_Q^k|^{1+\frac{2}{t+1}}\right)
\end{equation}
where $E_{OST}(\mathcal{T}_Q^k)$ denotes the edge set of the optimal Steiner tree for query $Q_k$.

\section{TABLE TRANSFORMATION LAYER} \label{sec:transform}
The table transformation layer (1) automatically generates the transformation code that enables table joins (Sec. ~\ref{sec:transform_exe}), and (2) executes these transformations efficiently through a graph-based parallel strategy (Sec. ~\ref{sec:parallel}).

\subsection{Transformation Execution}\label{sec:transform_exe}

For each edge $e_k^{ij} \in E_k$, \sys uses a transformation module $M_T$ to produce executable code to join the two columns in $T_k^i$ and $T_k^j$ specified by the LLM in the last later. Specifically, $M_T$ identifies noise that interferes with equi-joins between the two tables and generates transformation code to fix these inconsistencies, thereby enabling an equi-join. As a modular component, $M_T$ can be instantiated flexibly with various methods, such as DTT~\cite{dtt}, TabulaX~\cite{tabulax}, or our proposed ReAct-style LLM agent. At a high level, the advantage of our approach is that it enables the agent to autonomously decide both when to invoke transformation operations and which specific ones are appropriate. Before proceeding, we must first give formal definitions of the grammars for several required operations.


\noindent \textbf{Transformation Grammar.} As shown in Table~\ref{tab:operator_mapping}, we define 4 significant operators, organized into two functional categories: 

\noindent \underline{\textit{Validate ($\mathcal{O}_{val}$).}} 
Given the two columns corresponding to the tables, the \texttt{ValidateJoin} operator prompts the LLM to determine whether a join between them is feasible. To support this decision, we supply the join containment ratio as a quantitative signal:
$\text{Containment}(A, B) = \frac{|A \cap B|}{|A|}$
By comparing this ratio before and after each transformation, the agent checks whether the latest operation resolves inconsistencies. The procedure terminates and returns ``joined'' if the LLM concludes that no additional improvements can be made (for instance, when the ratio reaches $1$), or when the maximum number of iterations is reached. Otherwise, the agent proceeds by applying further suitable operators from $\mathcal{O}_{trans}$.

\noindent \underline{\textit{Transform ($\mathcal{O}_{trans}$).}}
We formally define the set of transformation operators used to support table joins as $\mathcal{O}_{trans} = \{ \texttt{Normalize}, \texttt{Split}, \\ \texttt{Pivot} \}$. 
In particular, the \texttt{Normalize} operator allows the agent to generate distinct transformation functions for different formats within a single column (e.g., standardizing ISO strings and Unix timestamps together). The \texttt{Split} operator breaks a string column into two atomic columns based on a given delimiter (e.g., “\_”), which facilitates accurate matching when dealing with composite keys. Lastly, the \texttt{Pivot} operator restructures the table so that schemas are aligned for downstream join operations.

\noindent \textbf{Iterative ReAct Workflow.} To summarize, the agent follows an iterative ``Reason-Act-Observe'' loop to  join the tables. In each iteration:
(1) The agent analyzes whether the tables need transformations based on the observation from previous iterations as well as current joinability score. If so, it selects the appropriate operators from the grammar.
(2) The agent executes this code to apply specific transformations to the tables.
(3) The environment captures the execution results and feedbacks the updated state.
This loop persists until either the LLM determines as joinable or the maximum iteration limit is reached.
Notably, this workflow supports multi-step transformation chains. For example, the agent can first invoke \texttt{Pivot} to align the table's overall structure. It can then apply \texttt{Normalize} to resolve any remaining cell-level inconsistencies.

\begin{table}[t] 
\small
\centering
\caption{Transformation Grammar in \sys.}
\vspace{-1em}
\label{tab:operator_mapping}
\renewcommand{\arraystretch}{1.3} 
\resizebox{\columnwidth}{!}{
\begin{tabular}{|c|l|l|} 
\hline
\textbf{Category} & \textbf{Operator} & \textbf{Operator parameters} \\ \hline
Validate & ValidateJoin & source\_col, target\_col \\ \hline
\multirow{3}{*}{Transformation} 
  & Normalize~\cite{fan2024autoprep} & input\_col, target\_format \\ \cline{2-3} 
  & Split~\cite{pandas_split} & split\_col, delimiter \\ \cline{2-3} 
  & Pivot~\cite{pandas_pivot} & reshape\_type, target\_cols \\ \hline
\end{tabular}%
}
\end{table}

\subsection{Parallelization Strategy}~\label{sec:parallel}
We obtain the validated Steiner trees $\{\mathcal{G}_1, \mathcal{G}_2, ..., \mathcal{G}_M\}$ in Sec.~\ref{sec:graph}, where each $\mathcal{G}_k = (\T_k, E_k)$. These trees are merged to construct the composite graph $\mathcal{G}_v = \left(\mathcal{T}_v, E_v\right)$, where $\mathcal{T}_v = \bigcup_{k=1}^M \T_k$ and $E_v = \bigcup_{k=1}^M E_k$.
A straightforward approach is to input the entire graph $\mathcal{G}_v$ to the ReAct-style agent and ask it to generate transformation code for every edge in that graph at once.
However, this would yield suboptimal accuracy. The reason is that the tree structure of join relationships introduces complex sequential dependencies that make it difficult for LLMs to reason about all transformations coherently in a single pass. However, sequentially processing each edge would incur significant latency. 
To address this, we propose an iterative parallel processing strategy. In each iteration, we identify independent edges—those sharing no common nodes—within $E_v$ and submit them to LLMs for parallel transformation. This approach ensures that transformations for independent edges can be reasoned about concurrently without cognitive interference.
The process iterates until all edges in the join path have been processed, effectively balancing transformation quality with computational efficiency. 


\noindent\textbf{Maximizing Average Parallelism.}
We partition the edge set $E_v$ into $C$ disjoint batches $\{I_1, I_2,..., I_C\}$, where each batch $I_t, t\in[1, C]$ contains a set of independent edges that can be transformed in parallel.
Then, our goal is to maximize the average parallelism of batches, which is defined as the mean number of transformations executed per batch, i.e., $\frac{1}{C}\sum_{t=1}^{C}|I_t|$.

Formally, the optimization objective could be formulated as:

\begin{equation}\label{eqa:4}
\begin{aligned}
\max_{\{I_t\}} \quad & \frac{1}{C}\sum_{t=1}^{C}|I_t| \\
\text{s.t.} \quad 
&  \bigcup_{t=1}^{C} I_t = E_v, \\
& \forall\,  t \neq t':\ I_t \cap I_{t'} = \emptyset, \\
& \forall\, e_{ij}, e_{uv} \in I_t:\ 
\{T_i, T_j\} \cap \{T_u, T_v\} = \emptyset.
\end{aligned}
\end{equation}

\noindent The second constraint ensures that the edges do not overlap between batches. The third constraint ensures that no two edges in one batch share the same table.

\begin{algorithm}[t]
\caption{Parallel Transformation Algorithm}
\label{alg:table-transformation}
\DontPrintSemicolon
\KwIn{Validated composite graph
$\mathcal{G}_v = (\T_v, E_v)$.}
\KwOut{Final integrated join path $\mathcal{G}^*=(\T^*,E^*)$.}
     \textcolor{blue}{\tcp{Step 1: Edge batching}}
$\mathrm{BatchList} \gets \mathrm{Batch}(\mathcal{G}_v)$;\nllabel{alg:alg2-1}

 $\T^* \gets \T_v$;\nllabel{alg:alg2-2}

 $E^* \gets \emptyset$;\nllabel{alg:alg2-3}

\textcolor{blue}{\tcp{Step 2: Execute batches in parallel}}
\For{each batch $I_t \in \mathrm{BatchList}$ }{\nllabel{alg:alg2-11}
    \ForPar{each edge $e_{ij} \in I_t$}{
              $\big(\hat{T}_i,\ \hat{T}_j,\ \hat{e}_{ij}\big)\gets \mathrm{ReAct\_Transform}\!\left(\ \T^*,\ e_{ij}\right)$;

              $\T^* \gets (\T^* \setminus \{T_i, T_j\}) \cup \{\hat{T}_i, \hat{T}_j\}$;

              $E^* \gets E^* \cup \{ \hat{e}_{ij} \}$; \nllabel{alg:alg2-15}
    }
}

 
    
    

      
        
        


    





\Return{$\mathcal{G}^*=(\T^*,E^*)$}; \nllabel{alg:alg2-18}
\normalsize
\end{algorithm}

\noindent\textbf{Parallel Transformation Algorithm.}
In Equation~\ref{eqa:4}, our goal is to choose as many independent edges as possible per batch, which corresponds to minimizing the total number of batches. This can be regarded as the classical edge-coloring problem ~\cite{konig1916graphen, vizing1965chromatic}, which assigns colors to edges such that no two adjacent edges share the same color, equivalent to our problem of finding independent edges. The goal is to minimize the number of colors used, which corresponds to the number of batches in our setting. \sys uses the typical edge coloring algorithm, i.e., Vizing~\cite{vizing1965chromatic} to identify a near-optimal list of batches (line~\ref{alg:alg2-1}, as shown in Algorithm~\ref{alg:table-transformation}). For each batch $I_t$, \sys executes all transformations in parallel using the ReAct reasoning loop, updates transformed tables, and records validated edges (line ~\ref{alg:alg2-2} - ~\ref{alg:alg2-15}). In the end, all transformed tables and validated edges form the final integrated join path (line ~\ref{alg:alg2-18}).

Next, we provide an example to demonstrate how \sys organizes transformations across multiple queries into parallel batches.

\begin{figure}[t]
    \centering
    \includegraphics[width=1\linewidth]{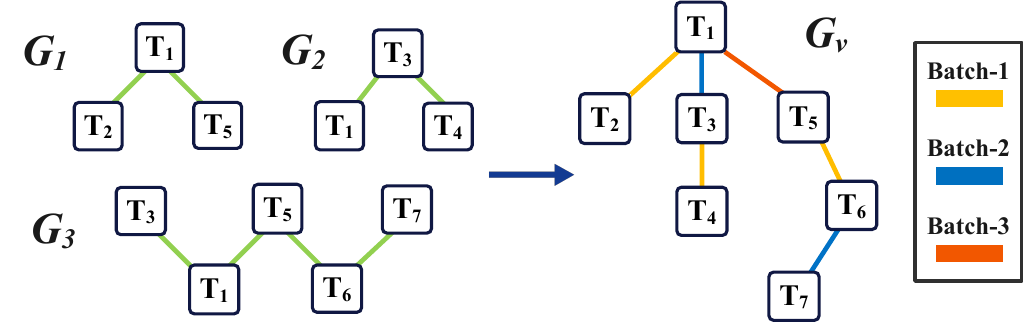}
    \vspace{-1.5em}
    \caption{Parallelization Strategy}
    \vspace{-1em}
    \label{fig:5.2-parallel}
\end{figure}

\begin{example}
As shown in Figure~\ref{fig:5.2-parallel}, we merge the Steiner tree $\mathcal{G}_1, \mathcal{G}_2, \mathcal{G}_3$ into the composite graph $\mathcal{G}_v$. We then apply the Vizing algorithm to color the edges of $\mathcal{G}_v$, which produces three batches of independent edges, i.e., $\text{Batch-1} =\{e_{12},e_{34},e_{56}\}$, $\text{Batch-2}=\{e_{13},e_{67}\}$, and $\text{Batch-3}=\{e_{15}\}$. Each batch runs simultaneously.

\end{example}

\section{Evaluation}~\label{sec:experiments}
In this section, we first detail our benchmark construction process (Sec. ~\ref{sec:exp_settings}), followed by extensive experiments to compare \sys against 6 baselines in both effectiveness and efficiency, focusing on answering the following research questions (RQs):

\noindent \textbf{RQ1} (Sec. ~\ref{sec:mainexp}): How does \sys perform in discovering the correct join paths compared to baseline approaches?

\noindent \textbf{RQ2} (Sec. ~\ref{sec:nl2sql}): Can \sys improve the accuracy of end-to-end SQL execution when combined with an Text-to-SQL system?

\noindent \textbf{RQ3} (Sec. ~\ref{sec:scalability}): How do the execution time and token cost of \sys scale as the total number of tables in the data lake increases? 

\noindent \textbf{RQ4} (Sec. ~\ref{sec:firstlayer}): How effective is the two-stage architecture in the table identification layer at retrieving relevant tables, and what is the impact of different PLM backbones? 

\noindent \textbf{RQ5} (Sec. ~\ref{sec:secondlayer}): How does the graph-based validation layer perform in estimating joinability and balancing accuracy with cost? 

\noindent \textbf{RQ6} (Sec. ~\ref{sec:thirdlayer}): How effective is the ReAct-style method in generating accurate table transformations, and how does the graph-based parallelization strategy improve execution efficiency?

\noindent \textbf{RQ7} (Sec. ~\ref{sec:error}): What are the main LLM error types in join validation and transformation generation, and what causes these failures in complex data lake settings?

\noindent \textbf{RQ8} (Sec. ~\ref{sec:ablation}): How do key hyperparameters (e.g., candidate table and training data ratios) impact \sys's performance?

\subsection{Experimental Settings}
\label{sec:exp_settings}

\begin{table*} [h!]
	\centering
	\caption{Statistics of Datasets.} 
        \vspace{-1em}
    {
		\resizebox{\linewidth}{!}{
            \begin{tabular}{c||ccccc}
		      \toprule
			\textbf{Dataset} & \textbf{\#-Tables} & \textbf{\#-Queries} & \textbf{\#-Max/Min/Avg. Column}& \textbf{\#--Max/Min/Avg. Row} & \textbf{Avg. \#-Tables/Join Relations Per Query}     \\ \hline
                \texttt{Ad}~\cite{EcoTableGitHub_ad}                & 69  & 126 & 86 / 3 / 19.51 &2630 / 1 / 99.51 & 4.65 / 3.65 \\ 
                \texttt{Engagement}~\cite{EcoTableGitHub_engagement}        & 56 & 38 & 48 / 3 / 14.11 & 100 / 1 / 20.32 &  3.76 / 2.76 \\
                \texttt{Business}~\cite{EcoTableGitHub_business}    & 90 & 28 & 389 / 2 / 48.66 &4033 / 1 / 100.48 & 5.33 / 4.33    \\
                \texttt{Platform}~\cite{EcoTableGitHub_platform}          & 35  & 24  & 15 / 3 / 7.43 &200 / 1 / 38.29 & 3.67 / 2.67    \\
                \texttt{NYC Data Lake}~\cite{EcoTableGitHub_NYC}          &  1214 & 800  &  219 / 1 / 19.67& 480 /  1 / 183.58&   5.04 / 4.14\\

                \bottomrule
		\end{tabular}
        }
	}
        \vspace{-1em}
	\label{tbl:dataset}
\end{table*}

\begin{figure*}[t]
\centering
\includegraphics[width=0.9\linewidth]{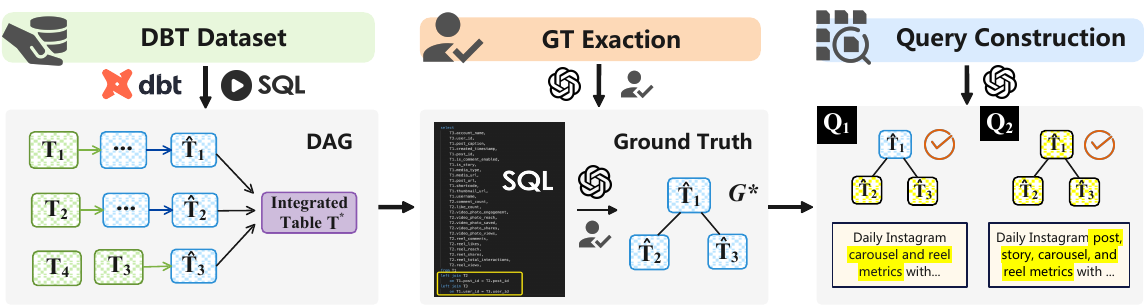}

\caption{DBT Benchmark Datasets Construction.}
\vspace{-1em}
\label{fig:dataset}
\end{figure*}

\noindent {\textbf{Datasets.}}
We evaluate the performance of \sys across five benchmark datasets, categorized two groups: four real-world industrial benchmarks and a large-scale synthetic data lake. The statistics of these datasets are summarized in Table~\ref{tbl:dataset}.

\noindent(i) \textbf{DBT Benchmarks}. We derive four datasets from well-known DBT datasets for table integration and transformation~\cite{dbt}. Dataset statistics are shown in Table~\ref{tbl:dataset}. Specifically, 
(1) \texttt{Ad} represents advertising analytics scenarios and includes 69 tables and 126 natural language queries. Each query typically involves information such as spend, impressions and clicks, with an average of 3.65 joins per query.
(2) \texttt{Engagement} captures user engagement and communication data, including email campaigns, in-product events, and support ticket records. It contains 56 tables and 38 queries, averaging 2.76 joins per query.
(3) \texttt{Business} consists of 90 tables and 28 queries. It integrates financial ledgers, orders, and sales management data, averaging 4.33 joins per query.
(4) \texttt{Platform} captures engineering and platform activities, such as code repositories, project management, and data pipeline logs. It includes 35 tables and 24 queries, averaging 2.67 joins per query.
Our benchmark features columns ranging from 2 to 389 for each table, with row counts varying significantly from 1 to 4,033, highlighting the complexity of integration.

In Figure~\ref{fig:dataset} (left), each DBT dataset includes a set of interrelated tables and SQL statements that serve as the basis for our ground truth generation and query construction. The datasets offer original raw tables in the data lake (green tables $T_1, T_2, T_3$), transformed tables derived from these raw tables ($\hat{T}_1, \hat{T}_2, \hat{T}_3$), and integrated tables $T^{*}$, created by joining transformed tables (e.g., $T^{*}$ from $\hat{T}_1, \hat{T}_2, \hat{T}_3$). Transformations and joins are defined by SQL statements, detailing how tables like $\hat{T}_1, \hat{T}_2, \hat{T}_3$ transform from $T_1, T_2, T_3$ and how they join to form $T^{*}$.

To construct our benchmark, overall, we start by translating the provided SQL statements to the ground-truth join paths. We then automatically generate the corresponding natural language queries that semantically align with these join paths.

\noindent \underline{\textit{Ground Truth Exaction.}}
As shown in Figure~\ref{fig:dataset} (middle), for each join SQL statement that corresponds to an integrated table $T^*$, we use LLMs to parse the SQL statement, identify the participating tables and join conditions, and then transform it into a ground-truth join path $\mathcal{G}^* = (\T^*, E^*)$, where $\T^* = \{\hat{T}_1, \hat{T}_2, \hat{T}_3\}$ and $E^*$ denote valid join relationships inferred from the SQL join conditions. To ensure correctness, we ask human experts to check and correct the join paths.

\noindent \underline{{\textit{Query Construction.}}} ~\label{sec:query construction}
According to the ground-truth join path, we design two categories of queries based on whether bridge tables are required during integration:

\noindent \textit{(1) Queries Requiring Bridge Tables}:  
For this type of query, we begin by randomly selecting multiple subsets of non-leaf nodes from the ground-truth join path to serve as bridge tables. Once the bridge tables are selected (i.e., $\hat{T}_1 \in \T^*$ in $Q_2$ from Figure ~\ref{fig:dataset} (right)), they are removed from the join path, and the remaining tables ($\hat{T}_2, \hat{T}_3 \in \T^*$) are examined to identify relevant columns that will serve as attributes in the user’s query. These selected columns are then fed to LLMs to construct the corresponding natural language query.


\noindent \textit{(2) Queries Not Requiring Bridge Tables}:  
Since this type of queries requires columns from all tables in $\T^*$ involved in the ground truth join path, we can simply translate the join SQL statement into a natural language query using LLMs.

Overall, 67.13\% of queries require bridge tables (145/216), whereas 32.87\% do not (71/216).

\noindent(ii) \textbf{NYC Data Lake}. To further test the scalability of \sys and its robustness, we introduce a large-scale benchmark constructed from NYC Open Data~\cite{nyc_open_data}. Unlike the DBT benchmarks, this dataset is  generated to simulate a massive, messy data lake environment. It comprises 1,214 tables and 800 queries, featuring different types of noise and a large search space designed to challenge table identification and join path discovery algorithms. Next, we illustrate the concrete process of table generation, join path generation and query construction.

\noindent \underline{\textit{Real-World Data Lake (Tables \& Join Paths) Construction.}}
At a high level, besides the larger size, the constructed data lake should have two key characteristics: (1) it should contain join ambiguity and schema noise to reflect the messy real-world scenario; and (2) it should include a number of tables that support multi-table join to answer our queries.

To this end, we first select a set of 100 tables with the largest number of columns from the original large NYC Open Data. Then, for each table $T$, we retrieve top-$k$ ($k$=3) tables from the original Data Lake that are the most joinable with $T$, which is measured by DeepJoin~\cite{deepjoin}, but the three tables might not successfully join with $T$. This step introduces both join ambiguity  and schema noise because the NYC Open Data naturally contains noisy information like incomplete values, unaligned schemes, fuzzy joins, etc. 

Second, we split each table with large number of columns to multiple sub-tables to form multi-join relationships. We first identify high-cardinality columns (many distinct values) and designate them as join keys, which connect nodes along the join path. All other columns are treated as non–join-key attributes and partitioned into disjoint subsets, each assigned to a different sub-table (node) on the path. The number and layout of sub-tables follow the join-path template defined in the DBT project. Adjacent nodes share at least one high-cardinality join-key column, which defines their join condition. 

To better reflect realistic settings, we conduct the following  steps over these sub-tables.
(1) Since real-world multi-table joins often produce sparse matches, to simulate this, we restrict the overlap to only 5\% of original tables and  assign the remaining 95\% of rows into different sub-tables. (2) We inject cell-level (e.g., case changes, whitespace removal, typos, or number variations) noise to 37\% join columns.  (3) We ask humans to identify 8\% columns to merge/split to produce column-level noise. (4) We also apply pivot/unpivot operations over 4.8\% columns. Overall, we construct over 521 joinable tables with explicit join-path ground truth.

Third, for all the tables considered above (denoted as the set $\mathcal{T}$), the most direct way to obtain ground truth would be to have humans examine every possible table pair to determine joinability, which is prohibitively costly. To mitigate this, for each table $T$ and each $T' \in \mathcal{T}$, we only send the pair to human evaluators if DeepJoin($T$, $T'$) > 0.8; otherwise, we treat the pair as non-joinable. Because some join paths have already been identified as ground truth in the second step above, the amount of required human annotation is substantially reduced.
  
\noindent \underline{\textit{Query Construction.} }
In this new benchmark, we create a total of 800 queries, consisting of 100 human-written queries and 700 LLM-generated queries.
For the 100 human-written queries, we strictly follow the annotation pipeline described in the ``Incorporation of human-written queries'' part above.
For the 700 LLM-generated queries, we use the same construction process as for the four original datasets. Specifically, this process involves three steps: (1) For each join path in the data lake, we first prompt LLMs to instantiate predefined SQL templates, thereby producing candidate SQL statements. (2) We then execute these SQL statements to validate their effectiveness, directly discarding any that returns excessive NULL values or duplicate rows. (3) Finally, we ask LLMs to convert the remaining valid SQLs into natural language questions. Human annotators subsequently revise these questions by hand to ensure natural fluency while strictly maintaining equivalence with the underlying SQL.

\noindent\textbf{Baselines.} We compare \sys with various baselines, including existing methods and variants of our own approach.


\noindent (1) \texttt{Auto-BI}~\cite{autobi} identifies a join path using deep learning models given multiple tables. 
For fair comparison, we apply our table identification layer to capture query-relevant tables. \texttt{Auto-BI} then optimizes the join path by targeting these relevant tables.


\noindent (2) \texttt{Auto-Prep}~\cite{auto-prep} jointly predicts transformations and join paths with dual deep-learning models. It also considers the entire data lake tables rather than focusing on tables specific to an NL query. To ensure a fair comparison, we utilize our table identification layer to identify the tables relevant to the query. Subsequently, \texttt{Auto-Prep} performs searches for transformations and joins within these relevant tables.

\noindent (3) \texttt{LLMs-Only-EcoTable (LOE)}  relies exclusively on LLMs for the entire pipeline, performing table identification, join path discovery, and transformation code generation without graph-based optimization.


\noindent (4) \texttt{Search-with-Greedy-Algorithm (SGA)} is a variation of our method that incrementally selects join edges with the highest joinability scores until all relevant tables are connected.

\noindent (5) \texttt{Without-ReAct-Reasoning (WRR)} is another variation that removes the ReAct-style loop, utilizing LLMs to produce the final transformation code directly in a single step. 

\noindent (6) \sys is our full-fledged solution. It uses DeBERTa-v3-Large for $M_S$, DeepJoin for $M_J$, and the ReAct-style agent for $M_T$.


\noindent\textbf{Metrics.} We evaluate the join path accuracy, success rate for Text-to-SQL, average cost and average execution time across all the datasets. 
(1) Join Path Accuracy. We take F1 score as the evaluation metric. Let $\mathcal{G}^{gt}_k=(\T^{gt}_k,E^{gt}_k)$ and $\mathcal{G}^{pred}_k=(\T^{pred}_k,E^{pred}_k)$ denote the ground-truth and predicted join paths for query $Q_k$, respectively. 
We employ LLMs to judge whether the two edges represent the same join relation. Then, the precision and recall are computed as
$\text{Pre}_k = \frac{|E^{pred}_k \cap E^{gt}_k|}{|E^{pred}_k|}$and $\text{Rec}_k = \frac{|E^{pred}_k \cap E^{gt}_k|}{|E^{gt}_k|}$ respectively. The F1 score for $Q_k$ is
$F1_k = \frac{2 \times \text{Pre}_k \times \text{Rec}_k}{\text{Pre}_k + \text{Rec}_k}$.
The overall accuracy is 
$\overline{F1} = \frac{1}{M}\sum_{k=1}^{M} F1_k$, where $M$ is the total number of queries.
(2) Success Rate for Text-to-SQL. We follow ~\cite{spider2} to adopt the success rate (SR) metric, which is commonly used in Text-to-SQL evaluation ~\cite{spider2} to measure the  fraction of queries that are correctly answered. 
(3) Transformation F1-score. Following~\cite{dtt}, we evaluate transformation quality by comparing the system-generated join cell pairs  with the ground-truth pairs . 
(4) Average Cost. For each query $Q_k$, let $C_k$ represent the total expense incurred by all LLM calls, which is calculated based on the API pricing ~\cite{anthropic-api, openai-api-pricing} calculated considering token usage. The average cost is determined as $\overline{C} = \frac{1}{M}\sum_{k=1}^{M} C_k$.
(5) Cost-Aware Utility score (CAU). This score combines the F1 score and the normalized cost efficiency using a weight factor $\alpha$ to balance their importance: $CAU = \alpha \cdot F1 + (1 - \alpha) \cdot \left(1 - \frac{Cost - Cost_{min}}{Cost_{max} - Cost_{min}}\right)$. We set $\alpha=0.8$ in this paper.
(6) Average Execution Time.
We evaluate the average execution time as $\overline{t} = \frac{1}{M}\sum_{k=1}^{M} t_k$, where $t_k$ denotes the end-to-end execution time.


\subsection{Comparison with Baselines (RQ1)}
\label{sec:mainexp}

\subsubsection{Join Path Accuracy w.r.t. All Queries.}\label{sec:exp_acc}
\begin{figure}[t]
\centering
\includegraphics[width=\linewidth]{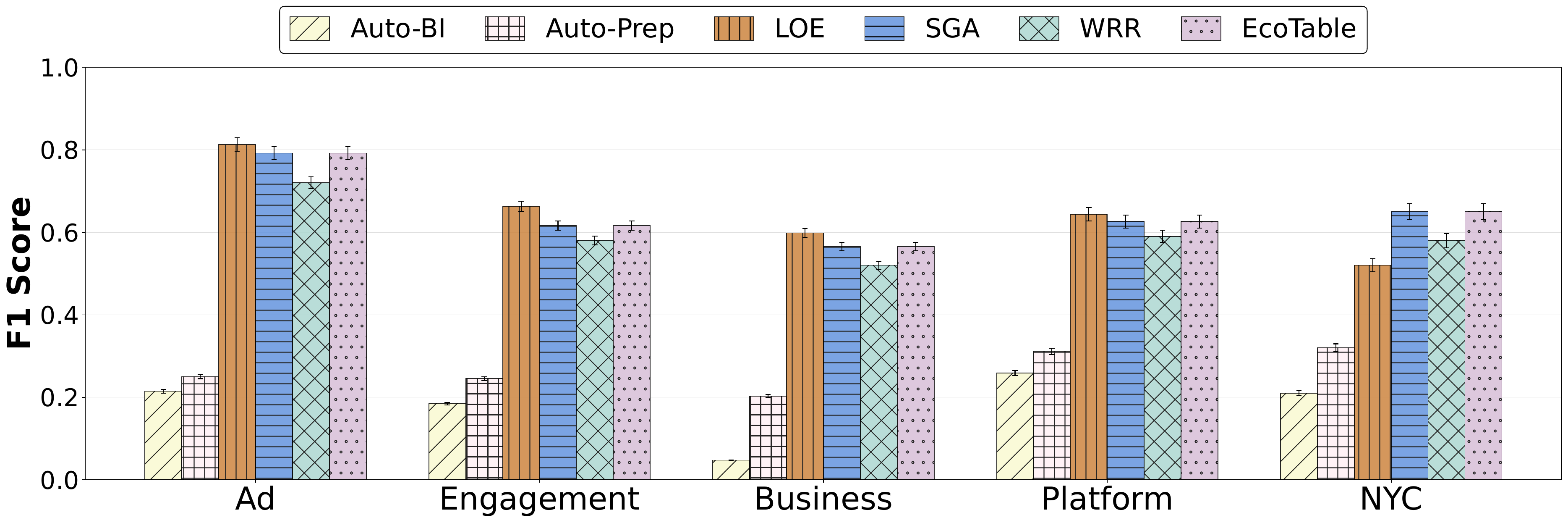}
\vspace{-2.5em}
\caption{Comparison of Join Path Accuracy (F1 Score).}
\vspace{-0.5em}
\label{fig:main_f1}
\end{figure}

Figure~\ref{fig:main_f1} reports the average F1 scores in join path accuracy with baselines. These methods are ranked as follows : \sys $\approx$ \texttt{LOE} $\approx$ \texttt{SGA} > \texttt{WRR} > \texttt{Auto-Prep} > \texttt{Auto-BI}.

\sys and \texttt{LOE} achieve the highest F1 scores among all baselines. \texttt{LOE} performs well because it leverages the powerful reasoning capabilities of LLMs to perform end-to-end join path discovery. \sys is competitive to \texttt{LOE} but incurs significantly lower costs. This is achieved by integrating a lightweight deep learning model with a Steiner tree formulation to identify a highly probable join path, which is then validated by LLMs solely on this path, thus reducing unnecessary LLMs calls while preserving high quality.
Empirically, \texttt{SGA} and \sys achieve similar F1 because \texttt{SGA} also uses LLMs to validate edges but incurs higher costs because of the suboptimal edge selection strategy.
Compared to \texttt{WRR}, \sys attains higher accuracy. because the ReAct mechanism effectively enables LLMs to detect and fix local reasoning errors during the transformation generation process.

\sys performs better than \texttt{Auto-BI}, for instance, on \texttt{Ad}, \sys achieves an F1 score of 79.2\%, significantly outperforming \texttt{Auto-BI} (21.5\%). This is due to (1) \texttt{Auto-BI} failing to consider table transformations, leading to a failure in joining two tables, and (2) \texttt{Auto-BI} depending solely on a deep model to estimate pairwise joinability. In contrast, \sys incorporates both table transformations and join path discovery using LLMs.

Notably, on  \texttt{NYC} dataset, \sys and \texttt{SGA} outperform \texttt{LOE}, which is different from the observations in other benchmarks. The F1-score of \texttt{LOE}  drops to  52\% because its initial schema linking step introduces many query-irrelevant tables that significantly reduce precision. In contrast, \sys maintains a high F1 score of 65\% by using a deep learning model to filter irrelevant tables, which ensures high recall while preventing LLM performance degradation due the large number of noise tables. Consequently, \sys proves more robust than end-to-end LLM methods when dealing with large-scale data lakes containing noisy data.



\subsubsection{Average Cost.}\label{sec:exp_cost}

\begin{figure}[t]
\centering
\includegraphics[width=\linewidth]{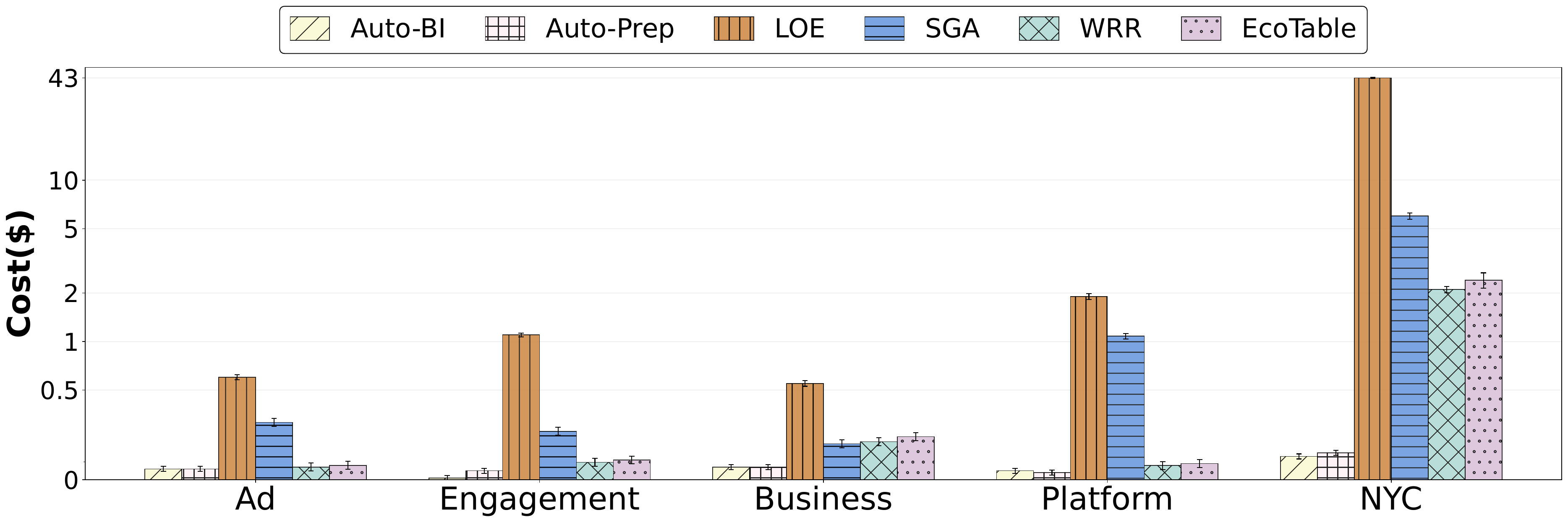}
\vspace{-2.5em}
\caption{Comparison of Token Cost per Query.}
\vspace{-1.2em}
\label{fig:main_cost}
\end{figure}

Figure~\ref{fig:main_cost} reports the average cost compared with baselines. \sys demonstrates a significant advantage in cost-efficiency compared to all baseline methods.

\sys has substantially lower cost than \texttt{LOE}. On average for all datasets, \sys reduces token cost by about $5.3\times$ compared to the greedy baseline \texttt{SGA}. This is because \sys leverages the Steiner-tree to achieve a global optimal edge selection, i.e., always identifying the most likely join paths for user-specified queries to validate. \sys  also remains cheaper than \texttt{LOE}, which relies on end-to-end LLM exploration of the entire join space. \texttt{Auto-BI} and \texttt{Auto-Prep} use fewer costs because they only use LLMs for relevant table identification. However, they are not accurate enough, as illustrated in Sec. ~\ref{sec:exp_acc}. 
\sys is slightly more expensive than \texttt{WRR} because the ReAct reasoning loop in \sys involves multiple iterative LLMs calls per transformation. However, the transformation code generated in a single LLM call often produces erroneous code due to a lack of execution feedback and step-by-step reasoning, resulting in a low F1 score of \texttt{WRR}.
On the \texttt{NYC} dataset, the cost of \sys rises because the join validation algorithm may exhaustively traverse edges to find a valid path in a large search space. To improve scalability, Section 6.4 presents a  variant of \sys with a sophisticated edge pruning strategy tailored for massive data lakes.



\begin{figure}[t]
\centering
\includegraphics[width=\linewidth]{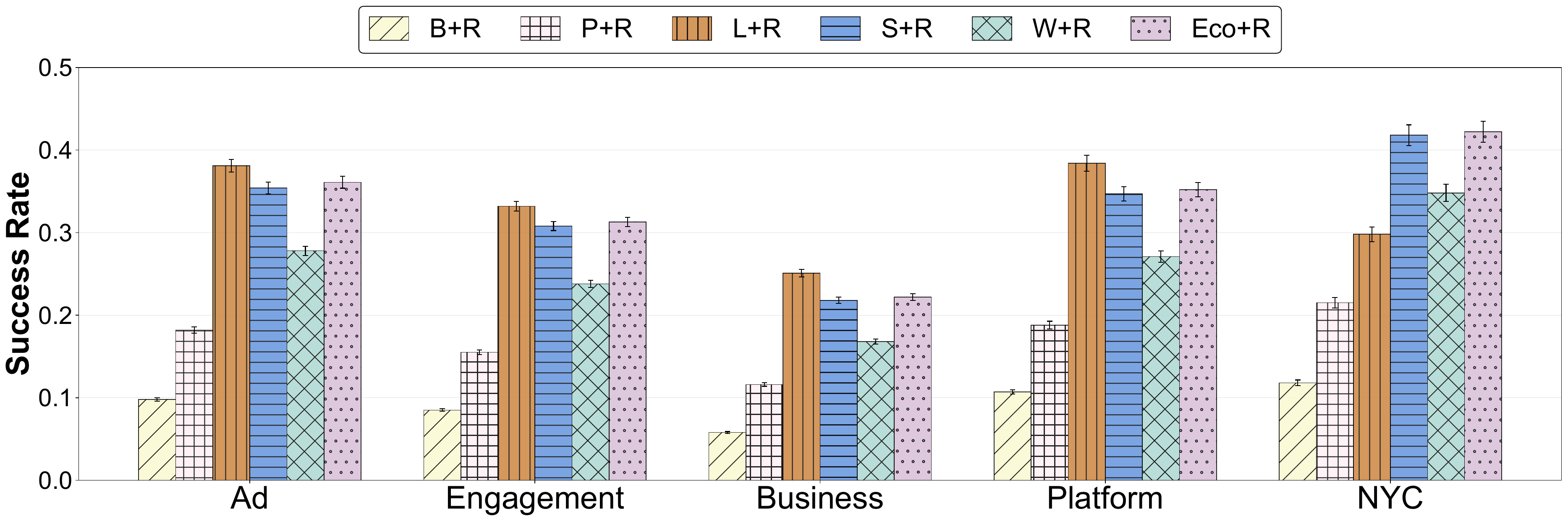}
\vspace{-2.5em}
\caption{End-to-end SQL Success Rate.}
\vspace{-0.5em}
\label{fig:nl2sql}
\end{figure}


\subsection{Integration with Text-to-SQL Systems (RQ2)}\label{sec:nl2sql}
We evaluate how our method affects the SQL execution accuracy when integrated with a Text-to-SQL model. We integrate our method with variant Text-to-SQL methods, including ReFoRCE ~\cite{reforce}, MAC-SQL ~\cite{macsql}, RSL-SQL ~\cite{RSLsql} and Spider-Agent ~\cite{spider2}. Due to space limitation, we only illustrate the results on the execution accuracy when integrated with ReFoRCE.
Furthermore, we compare \sys with the following baselines:







\begin{enumerate}[leftmargin=*]
\item \texttt{Auto-BI + ReFoRCE (B+R)} combines the join paths identified by \texttt{Auto-BI} with ReFoRCE.
\item \texttt{Auto-Prep + ReFoRCE (P+R)} merges the join paths detected by \texttt{Auto-Prep} with ReFoRCE.
\item \texttt{LOE + ReFoRCE (L+R)} integrates the join paths found by \texttt{LOE} with ReFoRCE.
\item \texttt{SGA + ReFoRCE (S+R)} incorporates the join paths generated by \texttt{SGA} with ReFoRCE.
\item \texttt{WRR + ReFoRCE (W+R)} leverages the join paths produced by \texttt{WRR} with ReFoRCE.
\item \texttt{\sys + ReFoRCE (Eco+R)} uses \sys to obtain verified join paths for ReFoRCE.
\end{enumerate}


In Figure ~\ref{fig:nl2sql}, our approach \texttt{Eco+R} achieves a competitive success rate compared to \texttt{S+R} and \texttt{L+R} across most datasets, but with significantly lower costs as illustrated in Sec.~\ref{sec:exp_cost}.
\texttt{Eco+R} outperforms both \texttt{B+R} and \texttt{P+R} because both methods adopt a deep learning model to estimate pairwise joinability without  LLMs validation, thus introducing incorrect join relationships to the Text-to-SQL model.
\texttt{Eco+R} outperforms \texttt{W+R} because its ReAct-based transformations effectively resolve data inconsistencies, providing clean tables that enable the Text-to-SQL model to generate executable and correct answers for the query.
Empirically, \texttt{L+R} and \texttt{Eco+R} achieve similar success rates on DBT benchmarks. However, on the NYC dataset, the performance of \texttt{L+R} drops because it introduces many query-irrelevant tables, which confuses the Text-to-SQL model and leads to incorrect join logic in the generated SQL.

\subsection{Scalability Evaluation. (RQ3)} \label{sec:scalability}
We evaluate how both the end-to-end runtime and token cost scale with  the total number of  tables in \texttt{NYC Data Lake}. Specifically, given the 800 queries, besides the tables necessary for answering these queries, we additionally add other tables into the data lake to incrementally expand the data lake size. The x-axis represents the number of added tables. Each query execution is repeated 10 times at each data scale.

\begin{figure}[t]
        \centering
        \includegraphics[width=1\linewidth]{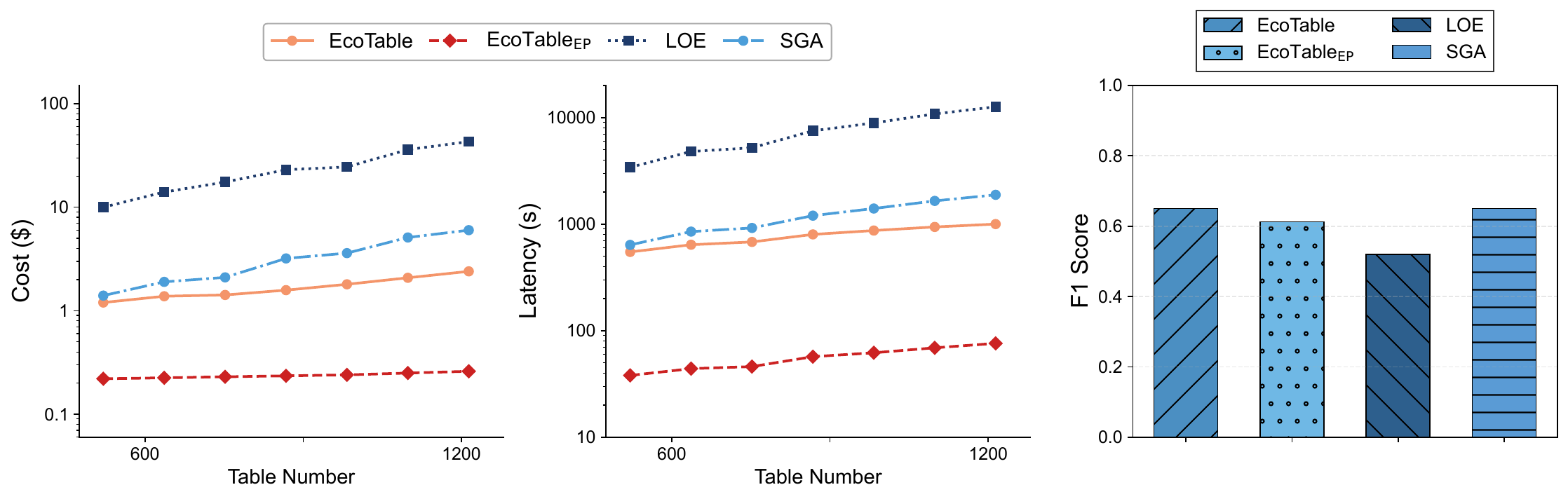}
        \vspace{-2.5em}
        \caption{Scalability Evaluation.}
        \vspace{-1em}
        \label{fig:scale}
    \end{figure}

 As illustrated in Figure~\ref{fig:scale} (a) and (b), we can observe that with the number of table increasing, \sys spends much more time and LLM costs than $\texttt{EcoTable}_{\text{EP}}$ because $\texttt{EcoTable}_{\text{EP}}$ prunes the original graph to a compact subgraph, and thus the search space is much reduced. $\texttt{EcoTable}_{\text{EP}}$ is also more efficienct than \texttt{LOE} and \texttt{SGA}, which further demonstrates its superior performance. Besides, we can observe that in Figure~\ref{fig:scale} (c), $\texttt{EcoTable}_{\text{EP}}$ achieves a similar performance with \sys, which does not sacrifice the accuracy because our pruned graph theoretically preserves the joinability property of the original large graph.

\subsection{Eval. of Table Identification Layer (RQ4)}
\label{sec:firstlayer}
We evaluate the table identification layer in \sys by focusing on two aspects: the impact of the underlying Pre-trained Language Model (PLM) and the effectiveness of our two-stage architecture.

\subsubsection{Impact of PLM Backbones.} \label{sec:layer1_plm} 
\begin{figure}[t]
\centering
\includegraphics[width=1\linewidth]{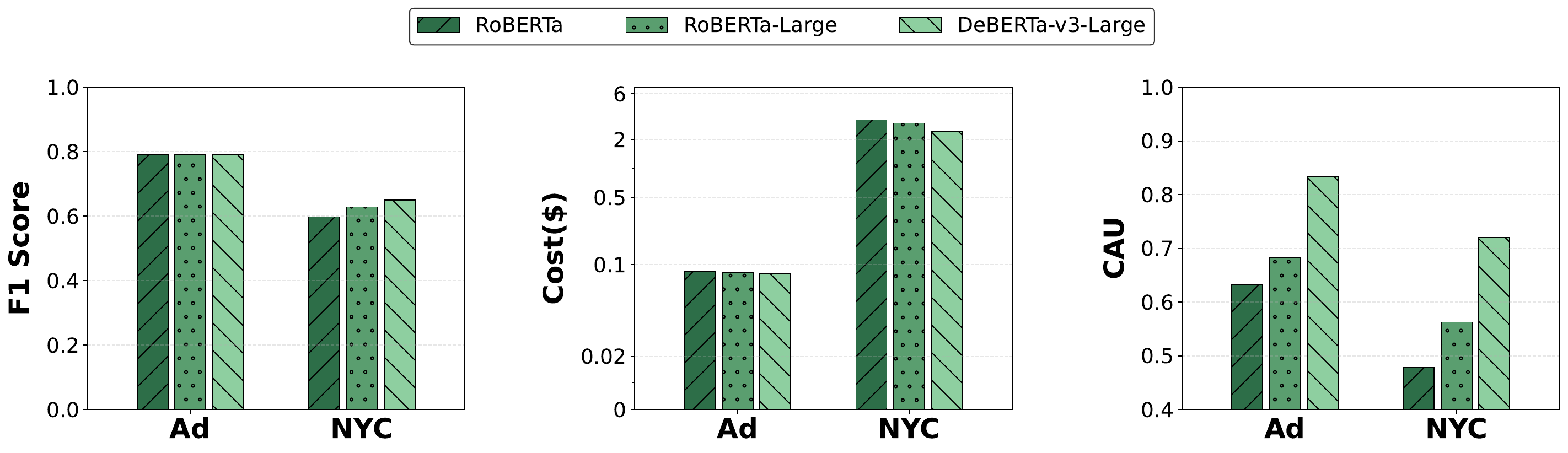}
\vspace{-2.5em}
\caption{Impact of PLM Backbones.}
\vspace{-1em}
\label{fig:layer1_plm}
\end{figure}

We compare multiple state-of-the-art models to evaluate how various PLMs influence teh result. In particular, we evaluate \texttt{RoBERTa}, \texttt{RoBERTa-large}, and \texttt{DeBERTa-v3-Large}.
As shown in Figure ~\ref{fig:layer1_plm}, 
These three models achieve similar accuracy and cost although \texttt{DeBERTa-v3-Large} uses a larger model, because we subsequently leverage the LLM to verify the results, which demonstrates that \sys is robust on the model used for schema linking.

\begin{figure}[t]
\centering
\renewcommand{\thefigure}{\arabic{figure}}
\includegraphics[width=1\linewidth]{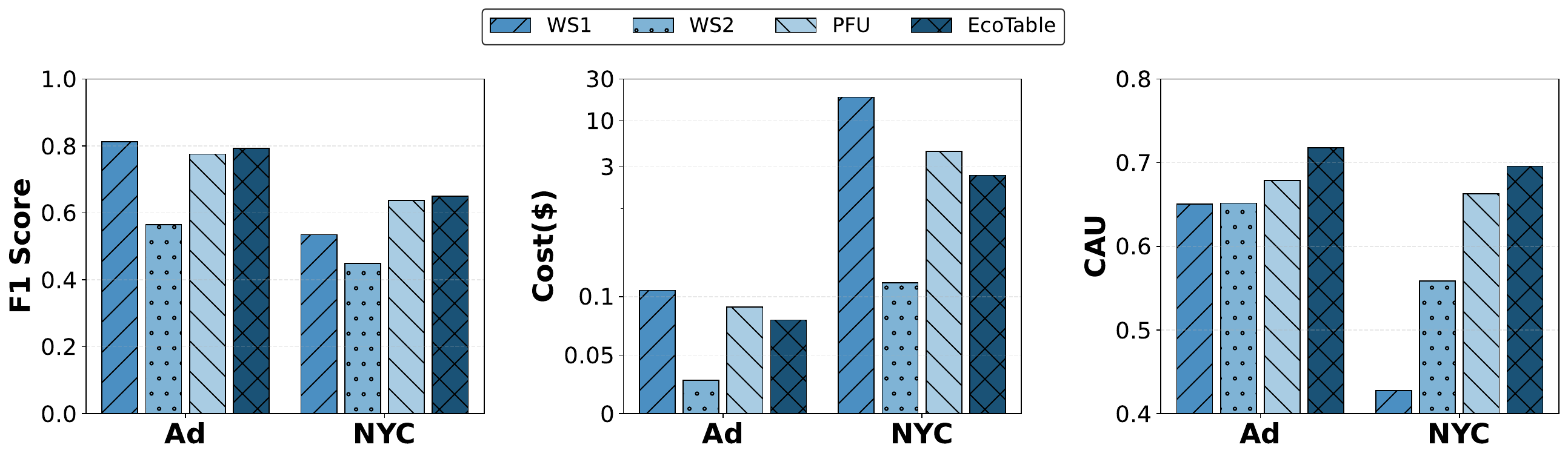}
\vspace{-2em}
\caption{Evaluation of Table Identification Layer.}
\vspace{-1em}
\label{fig:ablation_first}
\end{figure}

\begin{figure}[t]
    \centering
    \renewcommand{\thefigure}{\arabic{figure}}
    \includegraphics[width=1\linewidth]{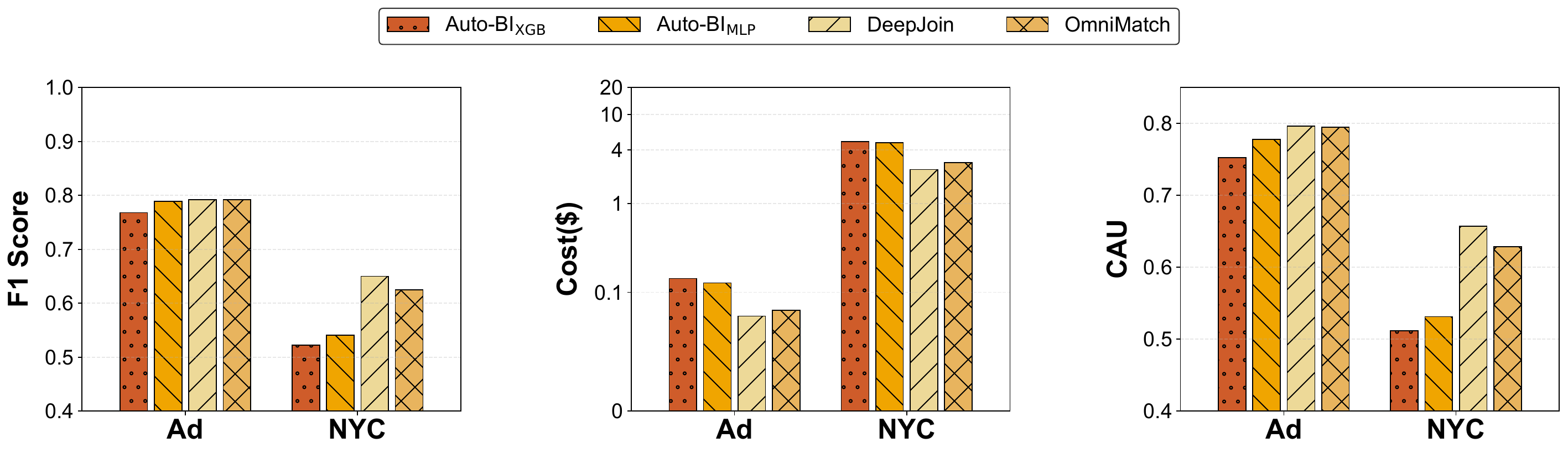}
    \vspace{-2.5em}
    \caption{Evaluation of Graph-based Validation Layer.}
    \vspace{-1em}
    \label{fig:validation_layer_cau}
\end{figure}  


\subsubsection{Effectiveness of Table Identification Layer.} \label{sec:exp_layer1_ablation}
We further evaluate the effectiveness of \sys's table identification layer by comparing it against 3  baselines:

\begin{enumerate}[leftmargin=*]
\item  \texttt{Without-Stage-1 (WS1)} removes the PLM-based coarse filtering in stage 1 from \sys and directly uses LLMs for schema linking.

\item  \texttt{Without-Stage-2 (WS2)} removes the LLMs refinement in stage 2 from \sys and keeps only the deep-model filtering.

\item  \texttt{Post-Filtering-Union (PFU)} uses LLMs to decompose a complex query into sub-queries, applies PLM-based filtering for each sub-query to find relevant tables, and then unions these sets before the LLM verification.

\end{enumerate}
Figure~\ref{fig:ablation_first} illustrates the Join Path F1-score for all methods.
\sys achieves the best performance across all baselines. \sys outperforms \texttt{WS1} because the deep-model filters irrelevant tables, ensuring high recall while preventing unnecessary LLM calls. 
Compared with $\texttt{WS2}$, \sys improves the average recall by 8\% across all the datasets with only a marginal increase in LLM invocations, as the LLM verification eliminates false positives from the deep-model filtering. 
%
%
\sys and \texttt{PFU} are competitive in Join Path F1 (about 61\%), but our method is cost-effective, because our method adopts a conservative filter strategy (by setting $k$ to a larger value, say 30) to discard tables that are unlikely to be relevant to a query.
%

\subsection{Eval. of Graph-based Validation Layer (RQ5)}
\label{sec:secondlayer}
We conduct a comprehensive evaluation of the Graph-based Validation Layer to validate the effectiveness of our joinability estimation module. Specifically, we compare \sys against the following methods:

\begin{enumerate}
[leftmargin=*]
    \item \texttt{Auto-BI$_{\text{MLP}}$}~\cite{autobi} takes a pair of columns, encodes them with 24 handcrafted features and feeds them into a four-layer MLP  to produce a probability.
    \item \texttt{Auto-BI$_{\text{XGB}}$}~\cite{autobi} relies on the same set of 24 manually engineered features, but performs classification using a Gradient Boosting Decision Tree estimators.
    \item \texttt{OmniMatch}~\cite{omnimatch} encodes each table column using statistical features and contextual embeddings learned with a Relational Graph Convolutional Network, which aggregates information from similar columns to refine their vectors and then computes distances between these vectors as a joinability probability.
    \item \texttt{DeepJoin}~\cite{deepjoin} converts table columns into textual sequences. It then applies a pre-trained MPNet model to obtain dense vector embeddings and assesses joinability by computing the cosine similarity between them.
\end{enumerate}

We can observe from Figure~\ref{fig:validation_layer_cau} that $\texttt{DeepJoin}$ and \texttt{OmniMatch} achieve a higher accuracy than $\texttt{Auto-BI}$ because the it leverages the pre-trained language model to compute the joinability, while Auto-BI just uses the hand-crafted features.
$\texttt{Auto-BI}$ also costs more because its low accuracy causes the LLM to check more edges.

\subsection{Eval. of Table Transformation Layer (RQ6)}
\label{sec:thirdlayer}

\subsubsection{Effectiveness of Transformation Generation.} 
\begin{figure}[t]
\centering
\includegraphics[width=1\linewidth]{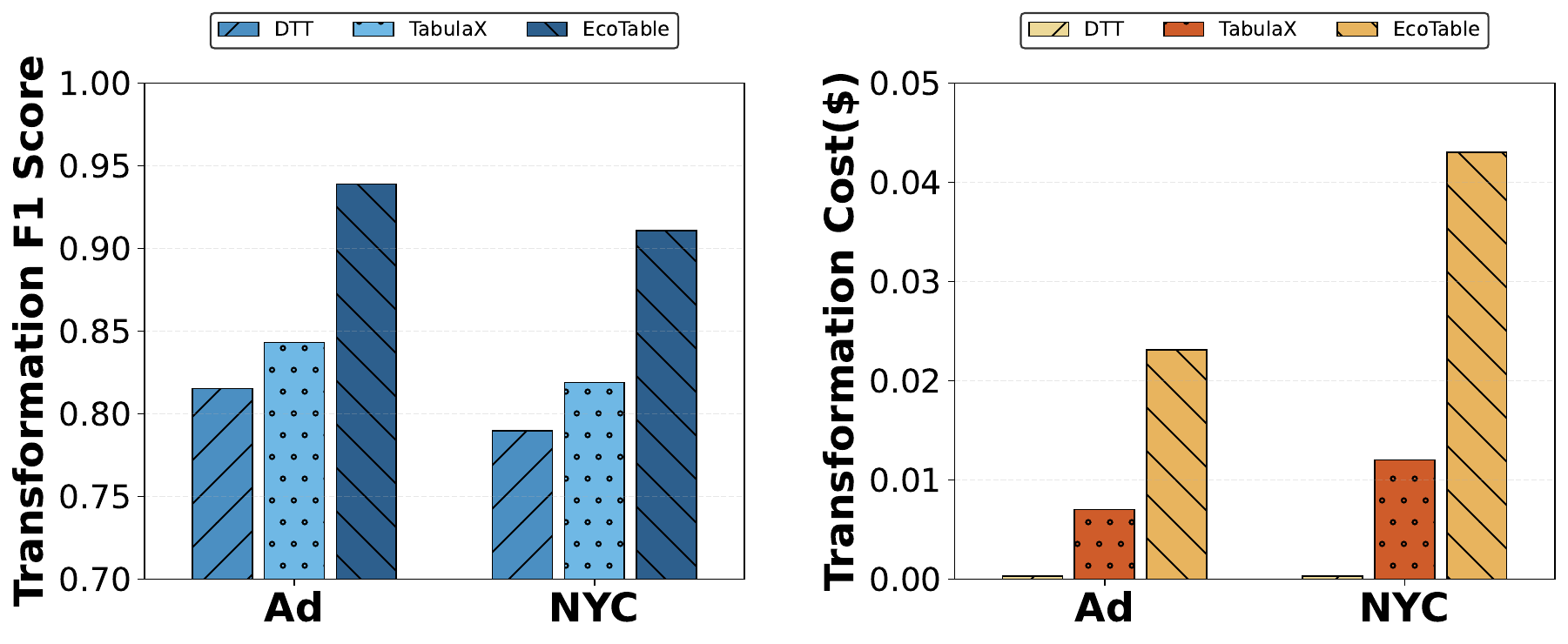}
\vspace{-2.5em}
\caption{Evaluation of Transformation Generation.}
\vspace{0em}
\label{fig:transformation_f1}
\end{figure}

To evaluate the accuracy of our ReAct-style method, we conduct a standalone evaluation comparing it against two automated baselines, i.e., \texttt{DTT} ~\cite{dtt} and  \texttt{TabulaX} ~\cite{tabulax} using the Transformation F1-score.
\begin{itemize}[leftmargin=*]
    \item \texttt{DTT} ~\cite{dtt}: A PLM-based baseline for table transformation. Given pairs of target columns that need to be aligned, \texttt{DTT} uses the original cell contents as input and directly produces the desired target values. The source columns are subsequently revised by mapping their cell values to these generated outputs, altering the tables so they satisfy equality-based join conditions.
    \item \texttt{TabulaX} ~\cite{tabulax}: An LLM-driven method for multi-class table transformations. The baseline first classifies each input column pair into one of four types: string-based, numerical, algorithmic, or general. Based on this, \texttt{TabulaX} uses LLMs to derive transformation rules, which are then applied to convert original cell values into target formats, enabling equality-based joins.
\end{itemize}
%
%
As shown in Figure~\ref{fig:transformation_f1}, \sys achieves a higher Transformation F1-score than both baselines because the ReAct-style reasoning dynamically composes atomic operators to resolve complex inconsistencies, whereas one-shot approaches like \texttt{DTT} and \texttt{TabulaX} struggle with unpredictable structural noise.

\subsubsection{Efficiency of Parallelization Strategy.} To evaluate the computational efficiency of our graph-based parallelization, we compare \sys against two execution baselines:
\begin{itemize}[leftmargin=*]
    
    \item \texttt{Sequential-Execution(SE)} executes transformations sequentially. For the graph $\mathcal{G}_v$, it sequentially invokes the ReAct-style agent on each edge to transform the table pairs and update the graph.
    \item \texttt{Composite-Graph (CG)} feeds the complete graph $\mathcal{G}_v$ to the React-style agent and requests it to produce transformation code for all edges in the graph simultaneously.
   
\end{itemize}

\begin{figure}[t]
\centering
\includegraphics[width=1\linewidth]{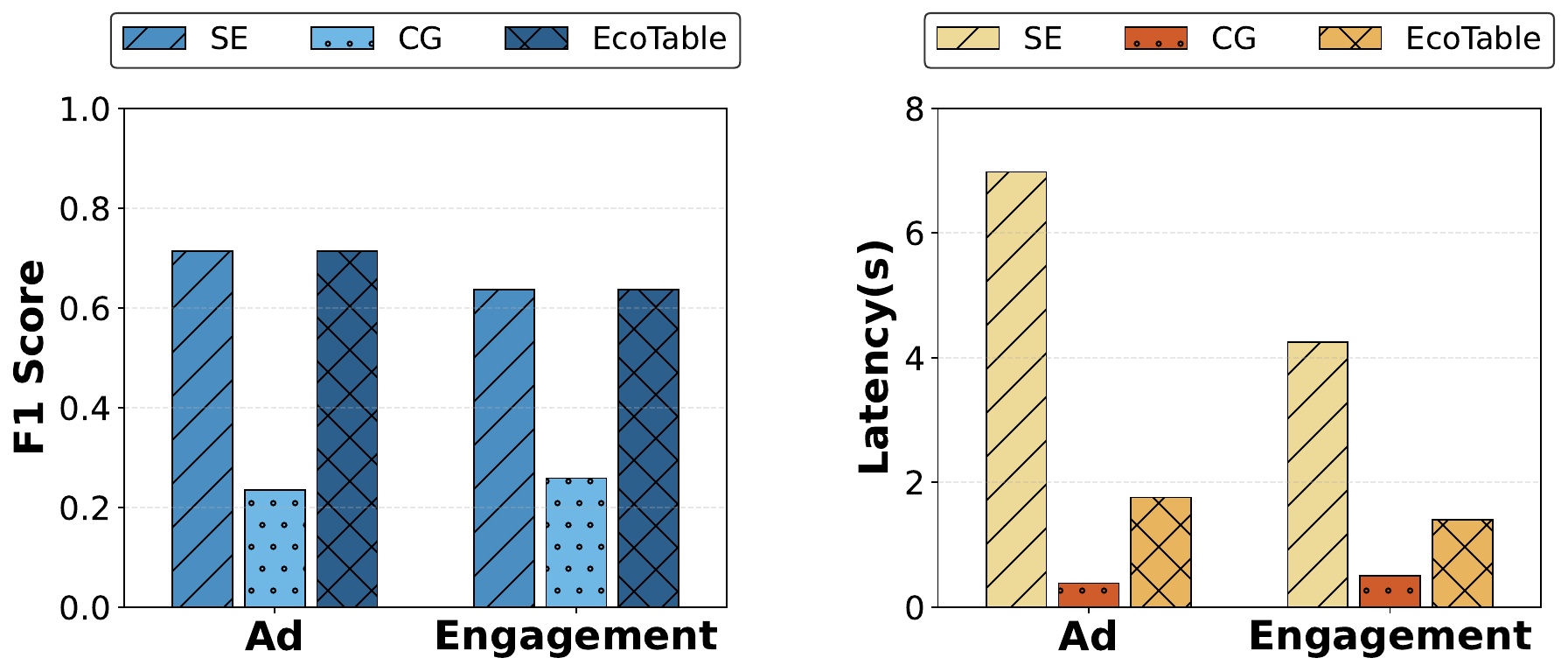}
\vspace{-2.5em}
\caption{Evaluation of Parallel Execution.}
\vspace{-1em}
\label{fig:parallel_execution}
\end{figure}

In Figure ~\ref{fig:parallel_execution}, our method achieves an average execution time of 6.29 seconds per query without compromising accuracy, making it 3.5 times more efficient than \texttt{SE}. This is because our parallel scheduling mechanism enables the concurrent execution of independent join validations and transformations, thereby optimizing resource utilization.
Although \texttt{CG} is highly efficient, it exhibits lower join path accuracy than \sys because the complex dependencies among transformations across the join path lead to hallucinations in LLMs or incorrect code for a significant portion of the edges.



\subsection{Error Analysis (RQ7)} \label{sec:error}
To understand where and why the LLM fails, we manually reviewed 100 error cases and categorized the LLM error modes in \sys into the following three types:

\begin{enumerate}[leftmargin=*]
\item \textit{Semantics Errors}. The LLM fails to understand the underlying semantic meaning of the data due to the contextual ambiguity.

\item \textit{Domain Knowledge Errors}. The LLM lacks external, domain-specific knowledge to reason correctly. 
\item \textit{Syntax Errors}. The LLM generates transformation code that fails at runtime due to syntax mistakes. 
\end{enumerate}

As shown in Table~\ref{tab:error_dist}, we compare the error distributions w.r.t. GPT-4o, Qwen3Max, and Deepseek3.2. These three types of errors manifest differently in our two core tasks: 
in join validation, the focus is mainly on textual reasoning instead of code execution, so syntax errors are irrelevant. Thus, only the other two error modes are observed in this task.
\begin{enumerate}[leftmargin=*]
\item \textit{Semantics Errors}: for instance, consider an \texttt{Internal\_Part\_Number} in a factory table (i.e., an ID used within a company’s manufacturing or inventory system) and a \texttt{Supplier\_Part\_Number} in a vendor table (i.e., an ID used by an external supplier in their catalog). Since the column names are similar and share example values (e.g., ``PN-8832''), the LLM may treat them as directly joinable. In fact, they are IDs from separate, unrelated systems.
\item \textit{Domain Knowledge Errors}: for example, a data lake includes a numeric \texttt{Code} column (e.g., ``541511'') and a textual \texttt{Course\_Name} column (e.g., ``Computer Programming'') that actually refer to the same entity. However, the LLM may still reject this relationship. because it lacks the domain knowledge that ``541511'' corresponds to ``Computer Programming''.
\end{enumerate}

\begin{table}[t]
\centering
\renewcommand{\arraystretch}{1.2} 
\caption{The Percentage of Different LLM Error Modes (\%).}
\vspace{-1 em}
\label{tab:error_dist}
\resizebox{\columnwidth}{!}{%
\begin{tabular}{|l|l|c|c|c|}
\hline
\multirow{2}{*}{\textbf{Task}} & \multirow{2}{*}{\textbf{Error Mode}} & \multicolumn{3}{c|}{\textbf{Model}} \\
\cline{3-5} 
 & & \textbf{GPT-4o} & \textbf{Qwen3Max} & \textbf{DeepSeek3.2} \\
\hline
\multirow{2}{*}{\textbf{Join Validation}} 
 & Semantics & 68& 53& 45\\
\cline{2-5} 
 & Domain Knowledge & 32& 47& 55\\
\hline
\multirow{3}{*}{\textbf{Transformation.}} 
 & Semantics & 55& 70& 67\\
\cline{2-5}
 & Domain Knowledge & 32& 21& 25\\
\cline{2-5}
 & Syntax & 13& 9& 8\\
\hline
\end{tabular}%
}\vspace{-1 em}
\end{table}

In Transformation Generation, we observe  all the three error modes:
\begin{enumerate}[leftmargin=*]
    \item \textit{Semantics Errors}: for example, given  \texttt{Company\_Name} column (e.g., ``Alphabet Inc'') and \texttt{Vendor} column (e.g., ``Google Cloud LLC''), the LLM relies on its prior world knowledge to equate the subsidiary with the parent company. It uses a \texttt{Normalize} operator to map ``Google Cloud LLC'' directly to ``Alphabet Inc.''. However, this transformation is invalid because these are two distinct legal entities in corporate tax filings, leading to a logically incorrect aggregation of financial records.

    \item \textit{Domain Knowledge Errors}: for instance, Table A contains a 10-digit \texttt{BBL\_Code} (e.g., ``1000477501''), while Table B has separate \texttt{Borough}, \texttt{Block}, and \texttt{Lot} columns. After splitting the 10-digit code, the LLM must map the first digit (``1'') to ``Manhattan''. Without possessing this city-specific knowledge, the LLM cannot generate the correct transformation code to align the two tables.

    \item \textit{Syntax Errors}: for instance, consider a multi-step transformation where the LLM first performs a \texttt{Split} operator on a \texttt{Full\_Address} column. The execution engine automatically assigns default names to the resulting output columns (e.g., \texttt{Full\_Address\_part\_1}). However, in the subsequent step, the generated code mistakenly refers to a semantic column name that does not exist (e.g., \texttt{Address}) when applying the \texttt{Normalize} operator. This inconsistency between the column name used in the generated code and the actual intermediate table columns results in a runtime error.
\end{enumerate}

We can observe from the above table that the semantic errors occur the most frequently, mainly because (1) the number of ambiguous  values in data lake columns is naturally large; 
(2) LLMs  tend to conduct over-reasoning from general language patterns; and  (3) database joins require exact logical matches, but LLMs operate on probabilistic language patterns that favor broad similarity over strict logic. Domain knowledge errors happen less because LLMs have already memorized extensive knowledge during pre-training, so it is less frequent than the semantic errors. In addition, the number of syntax errors is the smallest because we have an explicit grammar definition and a ReAct-style reasoning process.

\subsection{Ablation Studies (RQ8)}
\label{sec:ablation}

\subsubsection{Ratio of Candidate Tables in Stage 1.} 
\begin{figure}[t]
\centering
\includegraphics[width=1\linewidth]{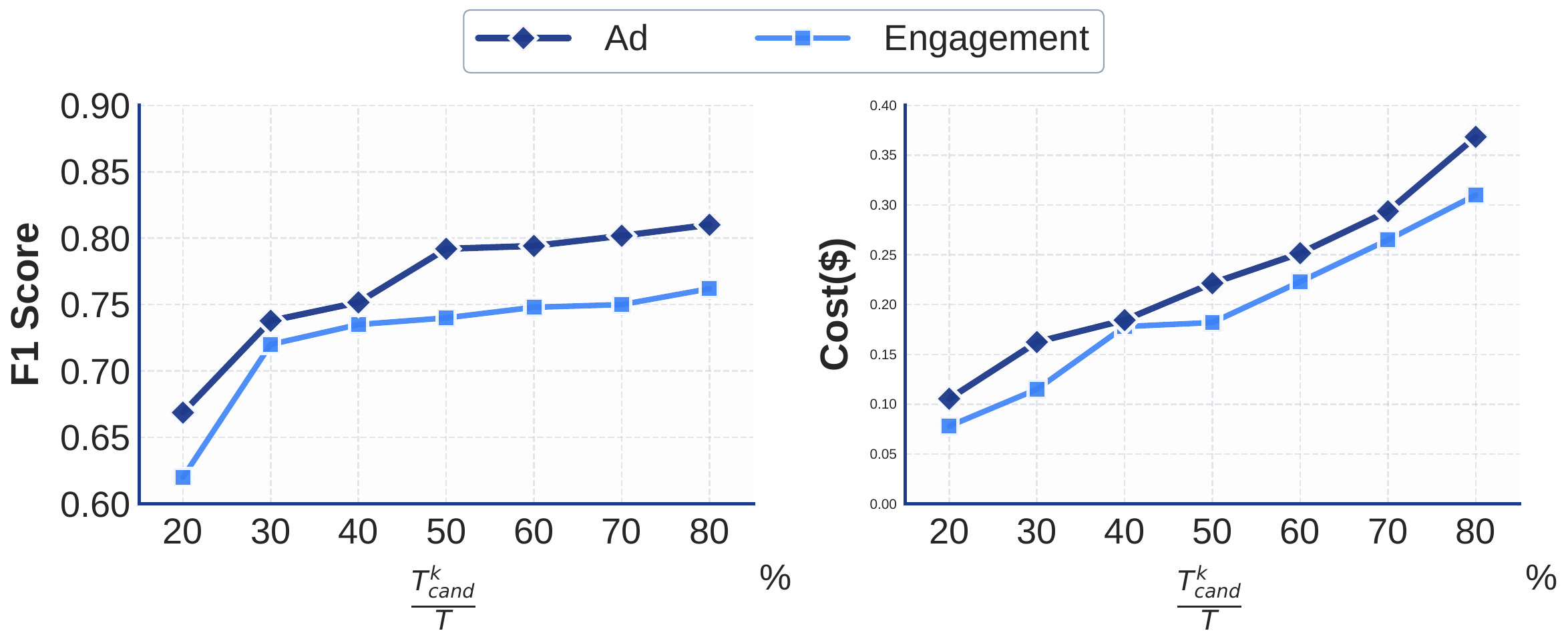}
\vspace{-2em}
\caption{Effect of the Ratio of Candidate Tables.}
\vspace{-1em}
\label{fig:topk}
\end{figure}

We evaluate the impact of the ratio of candidate tables $\frac{\T^k_{cand}}{\T}$. As shown in Figure~\ref{fig:topk}, we observe that a small ratio misses relevant tables, causing incomplete join paths and lower join path accuracy. While too large ratio leads to diminishing join path accuracy gains but significant cost overhead. Based on our empirical observation, setting $\frac{\T^k_{cand}}{\T}=0.3$ achieves the best balance between effectiveness and costs.

\subsubsection{Training Data Ratio.} 
\begin{figure}[t]
\centering
\includegraphics[width=1\linewidth]{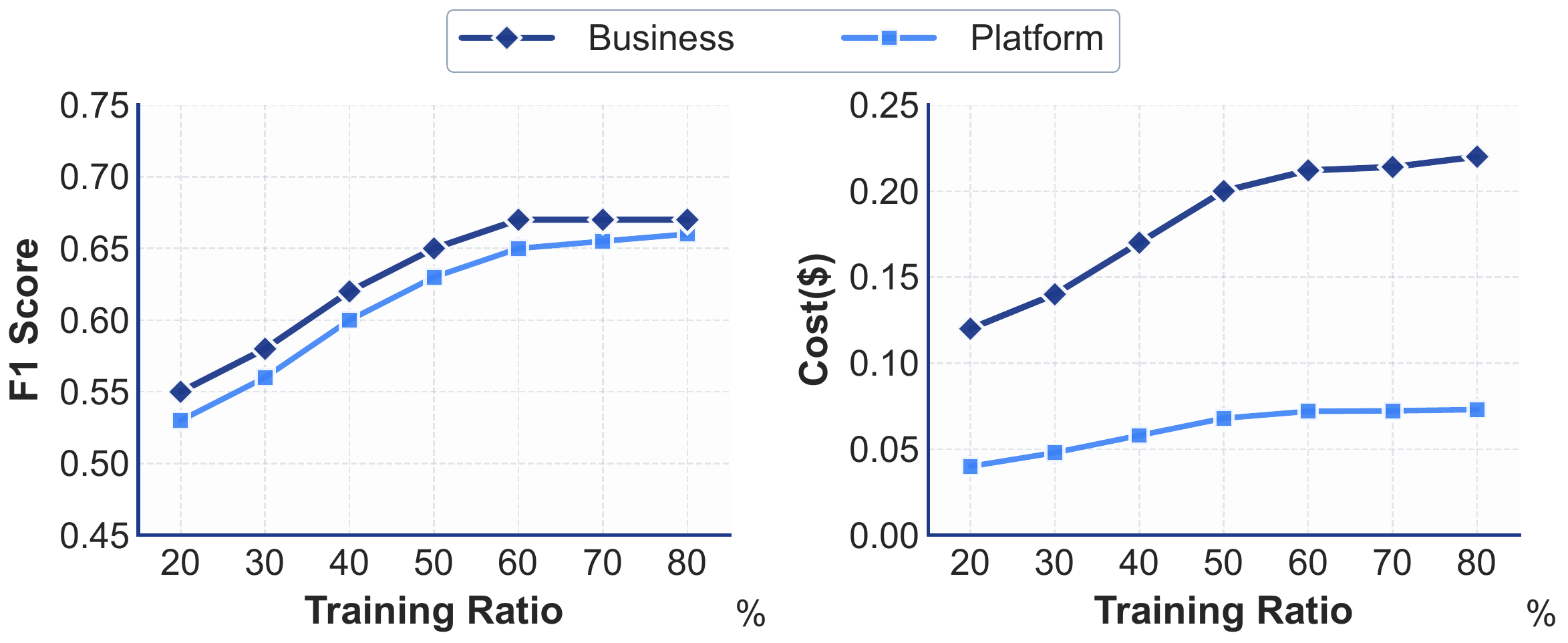}
\vspace{-2em}
\caption{Effect of the Ratio of Training Samples.}
\vspace{-1em}
\label{fig:train_ratio}
\end{figure}

We also evaluate how the ratio of training data used to fine-tune the deep model (i.e., RoBERTa ~\cite{roberta}) affects the performance of \sys. As shown in Figure ~\ref{fig:train_ratio}, with the ratio increases, the average F1 score improves until the ratio reaches 0.6. Therefore, a ratio around 0.6 appears to be suitable based on empirical observations.

\section{RELATED WORKS}
\label{sec:related-work}

\noindent\textbf{Text-to-SQL.}
Modern Text-to-SQL methods primarily follow two paradigms: fine-tuning, which adapts models to specific database schemas~\cite{resdsql,picard,codes} and LLM-based prompting, which leverages in-context reasoning without parameter updates~\cite{actsql,dinsql,mcssql}. As more challenging benchmarks such as Spider 2.0~\cite{spider2} and BIRD~\cite{bird} emerge, recent work increasingly focuses on strengthening reasoning. For instance, RSL-SQL~\cite{RSLsql} targets the semantic gap through schema linking, while MAC-SQL~\cite{macsql} and ReFoRCE~\cite{reforce} employ multi-agent interaction and iterative refinement to address ambiguous queries. Despite these advances, existing approaches assume that the databases that perfectly support user queries exist apriori, while our work targets lightly structured data lakes where no database schema is available beforehand.

\noindent\textbf{Joinable Table Discovery.} 
Discovering joinable tables in data lakes is essential for data integration. Its key component is to compute the joinability between two tables, which is also a significant step in our graph-based validation layer.
To be specific, early approaches such as Josie ~\cite{josie} and LSH Ensemble ~\cite{lsh} rely on set-level overlap between column values to measure the joinability. Pexeso ~\cite{pexeso} and DeepJoin ~\cite{deepjoin} leverage semantic embeddings from pre-trained language models to capture fuzzy joinability. \texttt{OmniMatch}~\cite{omnimatch} encodes each table column using statistical features and contextual embeddings learned with a Relational Graph Convolutional Network, which aggregates information from similar columns to refine their vectors and then computes distances between these vectors as a joinability probability.
Besides, schema matching methods~\cite{magneto, coma, adnev, llmmatcher} could also be utilized to compute the joinability based on the matching scores between two columns.
In contrast to these methods, which use a table to identify joinable tables in a data lake, we focus on processing natural language queries to integrate multiple tables within a data lake.

\noindent\textbf{Table Transformation.} Table transformation methods fall into two main categories: traditional methods~\cite{auto-join,auto-suggest,dtt} and LLM-based methods~\cite{tablegpt,tabulax}. For the former, early approaches like Auto-Join~\cite{auto-join} and Auto-Pipeline~\cite{auto-pipeline} rely on predefined operators and structural metadata to automate join discovery and pipeline construction, but are constrained by fixed operator sets. DTT~\cite{dtt} instead fine-tunes the PLM in an example-driven way to reformat data values. Among LLM-based methods, TableGPT~\cite{tablegpt} improves performance on table-related tasks via instruction tuning, while TabulaX~\cite{tabulax} classifies columns into four transformation types and applies chain-of-thought reasoning to generate interpretable functions. 
Unlike prior methods, we directly transform two given tables into a unified format that directly supports equi-joins for the specific join needs of each query.


\section{Conclusion} \label{sec:conclusion}
We introduced \texttt{EcoTable}, a cost-effective framework for integrating data lake tables via natural language queries. \sys uses a graph-based framework, combining lightweight models with LLMs to efficiently narrow the search space and minimize costly LLM calls through prioritized validation. Experiments show \sys achieves high integration accuracy but with lower costs.

\noindent \textbf{Limitation.} Our main focus is on queries with acyclic join graphs. We recognize that some complex queries might involve cyclic join graphs. Also,  there might be multiple join relationships between a table pair. These are valuable issues that we plan to explore in future work.

\bibliographystyle{ACM-Reference-Format}
\bibliography{sample-base}

\clearpage
\appendix

\section{Training Data Collection For $M_S$}
\label{app:train_data}
In the table identification layer, the deep learning model $M_S$ (i.e., RoBERTa) performs coarse-grained filtering to retrieve query-related tables. The training data consists of (NL query, table) pairs, each assigned a binary label $y \in \{0, 1\}$. This label denotes whether the table is relevant ($y=1$) or irrelevant ($y=0$) to the NL query.

We construct the initial training dataset through the following three steps: (1) we collect \texttt{DBT} projects and partition them into training and testing subsets; (2) we use LLMs to generate natural language queries from existing SQL in training sets, and then manually verify that each query is correctly aligned with its SQL; (3) we leverage \texttt{sqlparse} to identify tables referenced in the SQL and label them as relevant ($y=1$), while marking all remaining tables in the same data lake as irrelevant ($y=0$). To scale this process, we follow OmniSQL~\cite{omnisql} by generating a step-by-step Chain-of-Thought (CoT) explanation that describes the reasoning from the natural language query to the SQL code. This CoT serves as an automatic consistency check to guarantee semantic alignment between the query and SQL, eliminating manual verification. Using this procedure, we construct a large-scale training set with 200,000 labeled (query, table) pairs.

\section{Details of the Table Identification Layer}\label{app:prompt}

In the table identification layer, the LLM takes the original NL query and the coarse-grained candidate tables as input, and outputs the final verified table set. To achieve this, the LLM executes two guided operations: ``Query Understanding'' and ``Schema Reasoning and Verification''.
The detailed prompts used to guide the LLM in both operations are provided as follows.

\begin{tcolorbox}[
colback=white,
colframe=gray!75!black,
title=Prompt for Table Identification,
fonttitle=\bfseries,
breakable,
enhanced
]
\small
You are a data lake schema expert specializing in large-scale, open government data lakes. Your role is to identify, from a pool of candidate tables retrieved by a deep learning model, the minimal and
semantically correct set of tables required to answer a given natural language query.

You reason at the column level: a table is relevant if and only if its columns directly supply information requested by the query. In this data lake, tables may share naming prefixes and be organized by temporal scope or horizontal data partitioning; your reasoning must account for both aspects.

\smallskip
\noindent Step 1: Query Understanding \\
Given the following natural language query, decompose it into a structured list of atomic semantic components. Each component should represent one self-contained information need, such as:
\begin{itemize} [leftmargin=*]
\item A key entity or subject (e.g., "taxi trip", "building permit").
\item A measurable attribute or metric (e.g., "trip duration", "fare").
\item A filter or scoping condition (e.g., "yellow cab only").
\item A temporal specification (e.g., "in 2017", "FY 2018–2019").
\end{itemize}

Pay particular attention to temporal references. If the query mentions a specific year, fiscal year, or date range, record it explicitly, as it will be used to resolve temporal ambiguity among
candidate tables in the verification step.

\smallskip
\noindent\textbf{Input} \\
Query: \texttt{{query}}

\smallskip
\noindent\textbf{Output Format (JSON Only)}
\begin{verbatim}
{
    "temporal_scope": "<year / fiscal year / date range,
    or null>",
    "semantic_components": [
        "component 1: <description>",
        "component 2: <description>"
        ...
    ]
}
\end{verbatim}

\small
\noindent Step 2: Schema Reasoning and Alignment \\
You are given the original query, its decomposed semantic components (including temporal scope), and a ranked list of candidate tables from the data lake. Each table entry includes its name and a sample of its column names.

For each candidate table, reason step-by-step to determine whether the table directly contributes at least one column that satisfies a
semantic component of the query. Apply the following principles in order:

\begin{itemize} [leftmargin=*]
\item Column-Level Matching: A table is relevant only if its columns directly provide data requested by the query. Do not rely on table names alone.
\item Temporal Scope Consistency: If a temporal scope exists, discard tables whose names encode a different year or fiscal year, even if schemas match.
\item Partition-Complete Selection: For horizontally partitioned sub-tables (e.g., \texttt{records}, \texttt{records\_1}), examine and retain each variant independently if they are individually relevant.
\item Non-Redundancy: If two tables have identical schemas and partitions, retain only the one with broader coverage or the lower partition index.
\end{itemize}

\smallskip
\noindent\textbf{Input} \\
\texttt{Query: \{query\}} \\
\texttt{Temporal Scope: \{temporal\_scope\}} \\
\texttt{Semantic Components: \{semantic\_components\}} \\
\texttt{Candidate Tables: \{candidate\_tables\_with\_columns\}} \\

\smallskip
\noindent\textbf{Output Format (JSON Only)}
\begin{verbatim}
{
    "table_reasoning": {
        "<table_name>": {
          "matched_components": ["component i"],
          "matched_columns": ["col_a"],
          "temporal_consistent": true/false,
          "verdict": "keep/discard",
          "reason": "<one-sentence justification>"
        },
    ...
    },
    "selected_tables": ["table_1", "table_2", ...]
}
\end{verbatim}
\end{tcolorbox}

\section{Details of Pairwise Joinability Estimation}
\label{app:join-model}

We give details on the deep learning model $M_J(\cdot)$ introduced in Section~\ref{sec:validation}. Given a pair of columns $C_i$ from table $T_i$ and $C_j$ from table $T_j$, the model acts as a binary classifier to predict their joinability probability. Subsequently, this score is used to derive the edge weight $p(e_{ij})\in(0,1]$ between the two tables. In \sys, this model is a modular component instantiated using three methods: Auto-BI~\cite{autobi}, DeepJoin~\cite{deepjoin}, and OmniMatch~\cite{omnimatch}.

\noindent \textbf{Auto-BI~\cite{autobi}.}
Auto-BI represents each column pair $(C_i, C_j)$ with 24 features, categorized into \textit{Metadata-features} and \textit{Data-features}. We list these features, and precise definitions are in ~\cite{autobi}.

\noindent\underline{\textit{Metadata-features}}. 
These features capture structured information about column pairs, derived from table names and column names. In total, A total of 12 features derived from this metadata are utilized, including: Jaccard\_similarity, Jaccard\_containment, Edit\_distance, Embedding\_similarity, Token\_count, Char\_count, Col\_frequency, Col\_position, Col\_relative\_position, Unique\_col\_position, Table\_em-bedding, and Header\_jaccard. 

\noindent\underline{\textit{Data-features}}. 
13 features are extracted that relate to the data content of the columns. These features include Max\_containment, Min\_containment, Left\_containment, Right\_containment, Value\_di-stinct\_ratio, Range\_overlap, EMD\_score, Value\_length, Value\_type, Row\_cnt, Row\_ratio, Col\_ratio, and Cell\_ratio.

Given the limited information about the original model, we focus on two regression alternatives. We denote the MLP-based implementation as $\text{Auto-BI}_{\text{MLP}}$ and also introduce a tree-based ensemble variant, referred to as $\text{Auto-BI}_{\text{XGB}}$.

\begin{itemize} [leftmargin=*]
    
\item $\text{Auto-BI}_{\text{MLP}}$.
We employ a four-layer Multi-Layer Perceptron (MLP) as the regression model. Specifically, we concatenate the 24 features into a single vector and feed it into the MLP with hidden sizes $[256, 256, 128, 64]$. The output layer applies a Sigmoid function to produce a probability between 0 and 1.
For our experiments, we train the model by minimizing the binary cross-entropy loss.

\item $\text{Auto-BI}_{\text{XGB}}$.
We use $\text{Auto-BI}_{\text{XGB}}$, a variant that relies on Gradient Boosting Decision Trees implemented via \texttt{XGBClassifier}. In our experiments, we adopt this implementation for training and inference, using 100 decision trees.
\end{itemize}

\noindent \textbf{DeepJoin~\cite{deepjoin}.}
DeepJoin leverages a Pre-trained Language Model (PLM) to encode each column as a semantic vector, and then uses cosine similarity to assess whether two columns can be joined. Specifically, for each column of the pair, we present it as a text using the pattern: ``$table\_title$. $column\_name$ contains $n$ values ($max\_len$, $min\_len$, $avg\_len$): $cell_1$, $cell_2$, \dots, $cell_n$.''.
Here, $n$ denotes the number of distinct cell values in the column; $max\_len$, $min\_len$, $avg\_len$ are the maximum, minimum, and average numbers of words per cell, respectively; and $cell_i$ denotes the $i$-th distinct cell value selected based on frequency.
For this text, we then use MPNet ~\cite{mpnet} to compute column embeddings as input representations. For each pair, we denote the input as $(X_k, Y_k)$, where $k$ indexes the pairs and $X_k$, $Y_k$ are the embedding vectors for the columns in that pair.

To optimize this PLM, the Multiple Negative Ranking Loss is adopted in DeepJoin, which computes the negative log-likelihood of softmax-normalized scores:
\begin{equation}
    \mathcal{L} = -\frac{1}{N}\sum_{k=1}^{N} \left[ S(X_k, Y_k) - \log \sum_{l=1}^{N} \exp(S(X_k, Y_l)) \right]
\end{equation}
where $S(\cdot, \cdot)$ denotes the cosine similarity, and $N$ is the batch size.

\noindent \textbf{OmniMatch~\cite{omnimatch}.}
Unlike the methods above, OmniMatch builds a feature vector for each column by combining two components:  
1) representative statistics from that column, following the design in~\cite{sherlock}, and  
2) features learned by applying a GNN to a graph of semantically similar columns, so that information can be shared across similar columns.
In more detail, OmniMatch defines five pairwise column similarities and uses them to construct a global graph where nodes are columns and edges are weighted by similarity scores in (0, 1]. OmniMatch then applies an RGCN on the graph to aggregate information from columns that are similar to a given column. In this way, each column $C_i$ gets a final feature vector $h_i$ that captures these similarity relationships. 
Finally, to predict the joinability of $(C_i, C_j)$, we first compute the distance between their feature vectors $h_i$ and $h_j$, and then apply an exponential decay function $\exp(-\cdot)$, to map this distance to a joinability probability. In our experiments, we train the model parameters using a triplet margin loss.

\section{Details of LLM-based edge validation.}
\label{app:llm-validate}
In this section, we detail priority-aware edge validation and prompt construction.

 \noindent \underline{\textit{Priority-aware edge validation.}} We agree with the reviewer that if the LLM incorrectly invalidates a critical edge that is essential for table connectivity, other edges would be invalid.  To mitigate this, \sys first identifies {\it critical} edges as follows: 1) edges that appear in Steiner trees corresponding to multiple queries within the same iteration, and 2) edges whose deletion would prevent the construction of a new Steiner tree for a query in the subsequent iteration. For such a case, \sys makes three independent LLM calls to validate it and uses majority vote to make the final decision. This significantly reduces the impact of hallucinations and preserves reliable graph connectivity.

\noindent \underline{\textit{Table Serialization in Prompts.}} We use a two-phase serialization strategy that we believe addresses the reviewer's concern. In the first step, we would like to clarify that we do not involve all rows to avoid noise. Instead, we randomly sample about 50 rows respectively from a pair of tables. Then we feed them to the LLM and ask it to identify several  candidate column pairs that are likely to be joinable. Subsequently, we only consider the semantic relationships between these pairs in the second step. For each pair of tables, we provide all their distinct values for an in-depth comparison. The LLM then outputs a binary decision (1 for valid, 0 for invalid), indicating whether a join relationship exists.
This two-phase approach ensures that we collect sufficient overlapping values to avoid incorrect non-joinable decisions. At the same time, it keeps the prompts compact, thus reducing hallucinations and performance degradation. The prompt is detailed as follows:

\begin{tcolorbox}[
colback=white,
colframe=gray!75!black,
title=Prompt for Join Validation,
fonttitle=\bfseries,
breakable,
enhanced
]
\small
You are a rigorous data architect. Your task is to analyze two given data tables and determine whether there is a reasonable business connection between them.

\smallskip
\noindent\textbf{Background} \\
The goal of joining tables is to enrich the information used to answer a query. A "valid join" occurs when you combine complementary attributes from different perspectives within the same scope. For example, joining a school's basic info with its program details for the same year and region allows a query to access both dimensions.

Conversely, tables covering different time periods, regions, or categories are "parallel slices." Joining them does not enrich information; it corrupts the coherence of results by mixing data from incompatible scopes (e.g., mixing a 2019 catalog with a 2022 catalog).

\smallskip
\noindent\textbf{Workflow Explanation} \\
To ensure accuracy, we follow a two-step verification process:

Step 1: Semantic \& Scope Alignment \\
Examine the table names and sampled rows to identify if they share the same temporal, spatial, and categorical scope.
\begin{itemize} [leftmargin=*]
\item If scopes differ (e.g., \texttt{Sales\_2021} vs \texttt{Sales\_2023}), return an empty \texttt{selected\_pairs}.
\item If scopes align, identify column pairs that refer to the same specific attribute (e.g., \texttt{emp\_id} vs \texttt{staff\_code}). Avoid matching different metrics (e.g., \texttt{max\_speed} vs \texttt{avg\_speed}).
\end{itemize}

Step 2: Data-Level Validation \\
For the selected pairs, you will be provided with the full set of distinct values. You must determine if they are:
\begin{itemize}[leftmargin=*]
\item Joinable: Columns point to the same real-world entity and values match via exact equality or clear transformation.
\item Unjoinable: Irreconcilable differences in values or business context are revealed.
\end{itemize}

\smallskip
\noindent\textbf{Input Data (Step 1)} \\
\texttt{Table Names: \{TABLE\_NAMES\}} \\
\texttt{Table 1 Sample (50 rows): \{TABLE1\_SAMPLE\}} \\
\texttt{Table 2 Sample (50 rows): \{TABLE2\_SAMPLE\}} \\
\texttt{Candidate Pairs: \{CANDIDATE\_LIST\}}

\smallskip
\noindent\textbf{Output Format (Step 1 - JSON Only)}
\begin{verbatim}
{
  "same_scope": true/false,
  "selected_pairs": ["Table1.ColA <-> Table2.ColB"],
  "reason": "Business logic for semantic matching."
}
\end{verbatim}

\hrule \medskip

\noindent\textbf{Input Data (Step 2)} \\
\texttt{Selected Pairs: \{Table1.ColA <-> Table2.ColB\}} \\
\texttt{Table 1 Distinct Values: [val1, val2, ...]} \\
\texttt{Table 2 Distinct Values: [valA, valB, ...]}

\smallskip

\noindent\textbf{Output Format (Step 2 - JSON Only)}
\begin{verbatim}
{
  "can_join": "1/0",
  "alignment_sample": ["val1 <-> valA", "val2 <-> valB"],
  "reason": "Final decision based on value overlap."
}
\end{verbatim}
\end{tcolorbox}

\section{Comparison of \texttt{Non-LLM} v.s. \sys}
\label{app:non-llm-baseline}

In this section, we compare \sys with a new  baseline without involving LLMs, denoted as \texttt{Non-LLM}. This baseline replaces our LLM-based components with traditional deep learning methods: it employs \texttt{DeBERTa-v3-large} for table identification,  \texttt{DeepJoin} to predict join paths, and \texttt{DTT} to execute table transformations. 

\begin{figure}[t]
    \centering
    \includegraphics[width=1\linewidth]{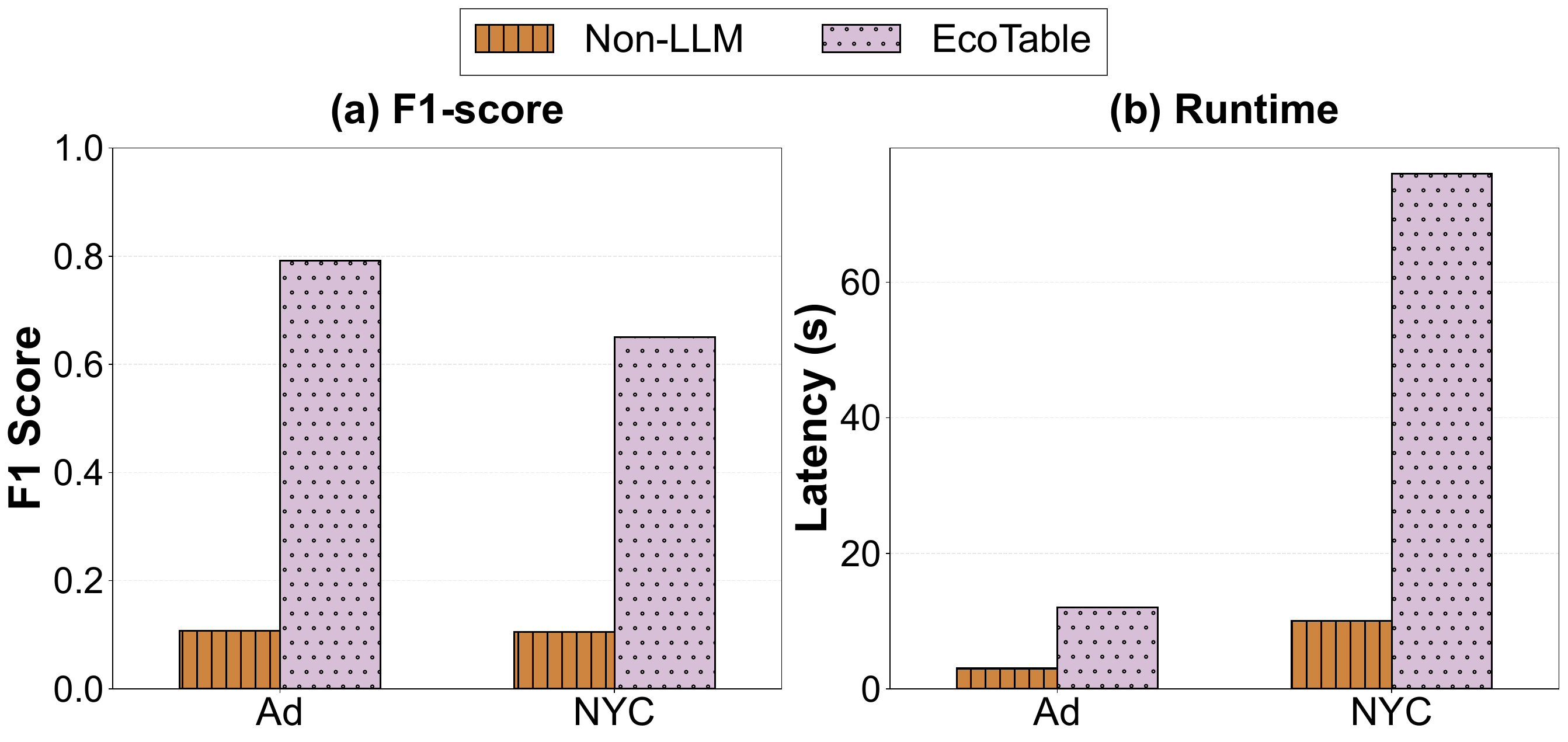}
    \vspace{-1em}
    \caption{Comparison of \texttt{Non-LLM} v.s. \sys.}
    \vspace{-1em}
    \label{fig:r1o12}
\end{figure}  

As shown in Figure~\ref{fig:r1o12}, \texttt{Non-LLM} is more efficient than \texttt{EcoTable}, as it completely avoids the high latency introduced by LLM inference. However,  \texttt{Non-LLM} is not effective  on complex integration tasks because its components depend on embeddings, which fails to handle the unpredictable noise present in real-world data lakes. In contrast, \sys achieves state-of-the-art performance by leveraging the adaptive reasoning and code-generation capabilities of LLMs. This enables \sys to flexibly identify required join paths and resolve complex data inconsistencies during runtime, substantially improving integration quality while only moderately increasing execution time.

\section{Statistics on Join Types and Transformations}
\label{dataset_stats}
we provide the join and transformation statistics in Table~\ref{tab:dataset_stats}.

\begin{table}[h]
\centering
\renewcommand{\arraystretch}{1.2} 
\caption{Statistics on Join Types and Transformations.}
\vspace{-1 em}
\label{tab:dataset_stats}
\resizebox{\columnwidth}{!}{%
\begin{tabular}{|l|c|c|c|}
\hline
\textbf{Dataset} & \textbf{\# Equi-joins} & \textbf{\# Fuzzy-joins} & \textbf{\# Total Transf.} \\
\hline
\textbf{Ad} & 65 & 17 & 17 \\
\hline
\textbf{Engagement} & 28 & 10 & 10 \\
\hline
\textbf{Business} & 13 & 7 & 7 \\
\hline
\textbf{Platform} & 11 & 8 & 8 \\
\hline
\textbf{NYC Data Lake} & 248& 204& 2440\\
\hline
\end{tabular}%
}
\end{table}

\section{Comparison of \texttt{MMQA} vs \sys.}
In this section, we incorporate the Multi-Table QA baseline: \texttt{MMQA} replaces our table identification layer with the retriever from MMQA. This method decomposes multi-hop questions into sub-questions and iteratively retrieves tables by jointly considering question-table and table-table relevance. 

\begin{figure}[t]
    \centering
    \includegraphics[width=1\linewidth]{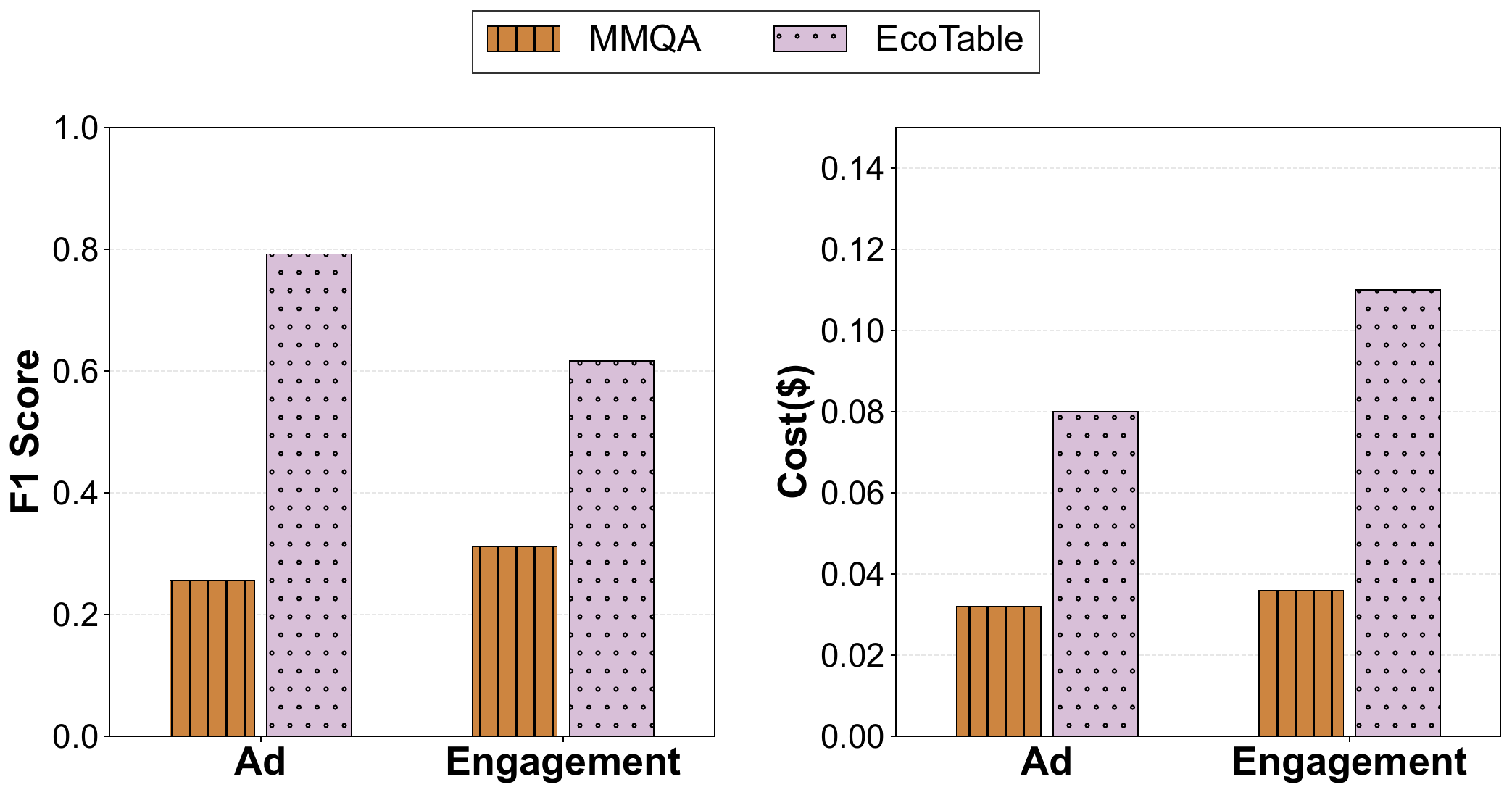}
    \vspace{-1em}
    \caption{Comparison of \texttt{MMQA} vs \sys.}
    \vspace{-1em}
    \label{fig:r4o3}
\end{figure}  

As illustrated in Figure~\ref{fig:r4o3}, \sys achieves a much higher Join Path F1-score (79.2\%) than \texttt{MMQA} (25.62\%) while incurring only a modest additional cost because \texttt{MMQA} uses a binary relevance score based on exact matches of column values across tables. Such a rigid criterion performs poorly in heterogeneous data lakes, where values are often noisy, inconsistent, or partially mismatched. If no exact overlap is found, its iterative retrieval terminates too early. In contrast, \sys is more resilient at filtering out irrelevant tables and preserving accuracy in large-scale settings like the NYC Data Lake. 
\texttt{MMQA} has a lower cost because it only uses LLMs to decompose queries.

\end{document}